\documentclass[%
 reprint,
superscriptaddress,
nofootinbib,
 amsmath,amssymb,
 aps,
prx,
]{revtex4-2}

\usepackage{graphicx}
\usepackage{dcolumn}
\usepackage{bm}
\usepackage{subfigure}
\usepackage{float}
\usepackage{placeins}
\usepackage{relsize}
\usepackage[normalem]{ulem}

\begin{document}

\preprint{APS/123-QED}

\title{Growth Laws and Universality in 2-TIPS: Microscopic and Coarse grained approach }

\author{Nayana Venkatareddy\textsuperscript{\textdaggerdbl}}
 \affiliation{Department of Physics, Indian Institute of Science, C. V. Raman Ave, Bengaluru 560012, India}
\author{Partha Sarathi Mondal\textsuperscript{\textdaggerdbl}}%
\affiliation{Department of Physics, Indian Institute of Technology (BHU) Varanasi, Uttar Pradesh 221005, India}%
\author{Jaydeep Mandal}
\affiliation{Department of Physics, Indian Institute of Science, C. V. Raman Ave, Bengaluru 560012, India}
\author{Shradha Mishra}
\email{smishra.phy@itbhu.ac.in}
\affiliation{Department of Physics, Indian Institute of Technology (BHU) Varanasi, Uttar Pradesh 221005, India}
\author{Prabal K. Maiti}
\email{maiti@iisc.ac.in}
\affiliation{Department of Physics, Indian Institute of Science, C. V. Raman Ave, Bengaluru 560012, India}

\date{\today}

\begin{abstract}
Two temperature induced phase separation(2-TIPS) is a phenomenon observed in mixtures of active and passive particles modeled by scalar activity where the temperature of the particle is proportional to its activity. The binary mixture of `hot' and `cold' particles phase separate when the relative temperature difference between hot and cold particles defined as activity \(\chi\) exceeds a density dependent critical value. The study of kinetics in 2-TIPS, a non-equilibrium phase separation, is of fundamental importance in statistical physics. In this paper, we investigate 2-TIPS kinetics using molecular dynamics (MD) and coarse-grained (CG) modeling in 3D and 2D. The coarse-grained model couples two passive Model B equations for hot and cold particles, with coupling terms emulating the energy transfer between them by raising the temperature of cold particles and lowering that of hot particles, a key observation from the MD simulations. MD simulations reveal that at high densities, phase separation begins immediately after the quench, forming bi-continuous domains rich in hot or cold particles, similar to spinodal decomposition in passive systems. These interconnected domains are also observed in the coarse-grained model for the mixture’s critical composition. Both MD and CG models show dynamic scaling of the correlation function, indicating self-similar domain growth. Regardless of dimensionality, both methods report algebraic growth in domain length with a growth exponent of 1/3, known as the Lifshitz-Slyozov exponent, widely observed in passive systems.  Our results demonstrate that the universality of phase separation kinetics observed in passive systems also extends to the non-equilibrium binary mixture undergoing 2-TIPS.
\end{abstract}

\maketitle
\def\thefootnote{\textdaggerdbl}\footnotetext{These authors contributed equally to this work}\def\thefootnote{\arabic{footnote}}

\section{Introduction\label{secI}}
Active matter \cite{marchetti2013hydrodynamics,bowick2022symmetry,forster2018hydrodynamic,ramaswamy2019active,toner2005hydrodynamics,romanczuk2012active,simha2002hydrodynamic} refers to systems whose constituents consume energy at a local scale to either self-propel or exert mechanical forces. These inherently non-equilibrium systems show fascinating collective behaviors like MIPS (Motility-Induced Phase Separation), dynamic self-regulation, and giant number fluctuations. Active matter is ubiquitous in biological systems \cite{needleman2017active,liu2021viscoelastic,banerjee2020actin,juelicher2007active} spanning length scales from sub-cellular cytoskeleton and colonies of bacteria to flocks of birds or fish schools. Artificially synthesized self-propelled particles, active liquid crystals, robotic swarms, and granular matter on vibrating plates have fascinating applications in pattern formation, liquid crystal display, and drug delivery \cite{palacci2014light,sun2022micro,duclos2020topological,doostmohammadi2018active,C9SM02552A,guix2018self}.\\
In most active matter models, activity is represented by a self-propulsion force, which is inherently vectorial in nature \cite{romanczuk2012active,PhysRevLett.126.188002,solon2015active,angelani2014first}. However, in binary mixtures where particles interact solely through volume exclusion, it has been observed that differences in the intrinsic properties of the two species can lead to phase separation. These differences can be modeled as variations in activity, such as in active-passive mixtures, or as disparities in transport properties like diffusion coefficients or effective temperatures. Such systems are classified as scalar active systems. These models have been applied to understand various biological phenomena, such as chromosome positioning within the nucleus \cite{ganai2014chromosome} and the behavior of molecular motors \cite{li2017double}. Further, scalar activity models have been adopted to various other fields like spin glasses \cite{dotsenko1995introduction} and heteropolymers \cite{pande2000heteropolymer}.\\
Scalar active systems have been studied both numerically and analytically. Analytical approaches have mainly concentrated on energy transfer between particles during two-particle interactions \cite{grosberg2018dissipation}, and the current in the non-equilibrium steady state of the system in the low-density limit \cite{grosberg2015nonequilibrium}. Joanny et al. \cite{PhysRevResearch.2.023200} derived an effective free energy and a set of equations for the concentration fields of two species in the low-density regime. Their analysis showed that the coarsening kinetics in the system could be described by the dynamics of a single conserved order parameter, predicting a growth law of $\frac{1}{3}$. Numerical simulations of the scalar active model in diverse soft matter systems, such as Lennard-Jones (LJ) disks, dumbbells, liquid crystals, chiral rods, and polymers \cite{chari2019scalar,PhysRevE.107.034607,PhysRevE.107.024701,D3SM00796K,chattopadhyay2021heating,chattopadhyay2024stability}, have revealed phase separation between hot and cold particles, known as Two-Temperature Induced Phase Separation (2-TIPS). In the case of LJ particles and dumbbells, the cold particles in the phase-separated system form a crystalline solid phase, while the hot particles exist in a gaseous phase. In system of soft repulsive spherocylinders (SRS), 2-TIPS leads to a transition from isotropic to nematic and crystalline order in the cold SRS as activity increases. \\
While most previous studies have focused on identifying different phases, Weber et al. \cite{weber2016binary} argued in favour of $\frac{1}{4}$ growth law in the low-density limit for a binary mixture of Brownian disks with high and low diffusivity, differing from the results of Joanny et al. \cite{PhysRevResearch.2.023200}. 
Moreover, the overall kinetics of domain growth, the role of dimensionality of the system, etc. remains largely unexplored. In this study, we aim to address these open questions by investigating the domain growth kinetics in a binary mixture of hot and cold species. To simplify the analysis, we consider a binary mixture of Lennard-Jones (LJ) particles that are identical in all properties except for their temperatures. Based on insights from previous studies, we constructed a minimal Coarse Grained (CG) model comprising two coupled model B equations and compared its predictions with Molecular Dynamics (MD) simulations of the microscopic system. \\
The study of the temporal evolution of a homogeneous binary mixture as it phase separates, when quenched into the phase-separated region, is of fundamental importance in statistical physics and is of great experimental interest. The phase-separation kinetics in passive systems \cite{puri2009kinetics,PhysRevE.77.011503,PhysRevE.85.031140,10.1063/1.472839} is well studied by theoretical and computational approaches, primarily due to the emergence of universality in dynamics irrespective of the nature of microscopic interactions. Recent investigations of phase separation kinetics in pure active matter \cite{D0SM01762K, PhysRevE.108.024609,PhysRevLett.131.068201,caporusso2024phase} systems reveal that some active systems share the universality classes with their passive counterparts. So we ask two important questions in our work: How does the binary mixture modeled by scalar activity, initially in a mixed state, reach its non-equilibrium steady state when quenched into the phase-separated region? What are the similarities and differences between phase-separation kinetics between 2-TIPS (a non-equilibrium phase separation) and passive systems? \\

 The MD simulations reveal that kinetics in 2-TIPS at high density, proceeds by the formation and growth of bi-continuous domains rich in either hot or cold particles in both 3D and 2D. Our calculations in 3D and 2D, show dynamic scaling of correlation function implying self-similar domain growth and power law behavior of domain length with growth exponent approximately equal to 1/3. The results obtained from CG model for the 2-TIPS  show a good agreement with the MD results regarding the nature of domain morphology, dynamic scaling as well as growth exponent. Rest of the paper is organized as follows: The simulation details of MD simulations and CG model are presented in section \ref{sec:model}, while the results are elaborated in section \ref{sec:res}. Finally, we summarize our results and discuss the possible future directions for our work in section \ref{secIV}. \\

\begin{figure*}[ht]
\centering

\subfigure[]{
  \includegraphics[width=0.23\textwidth]{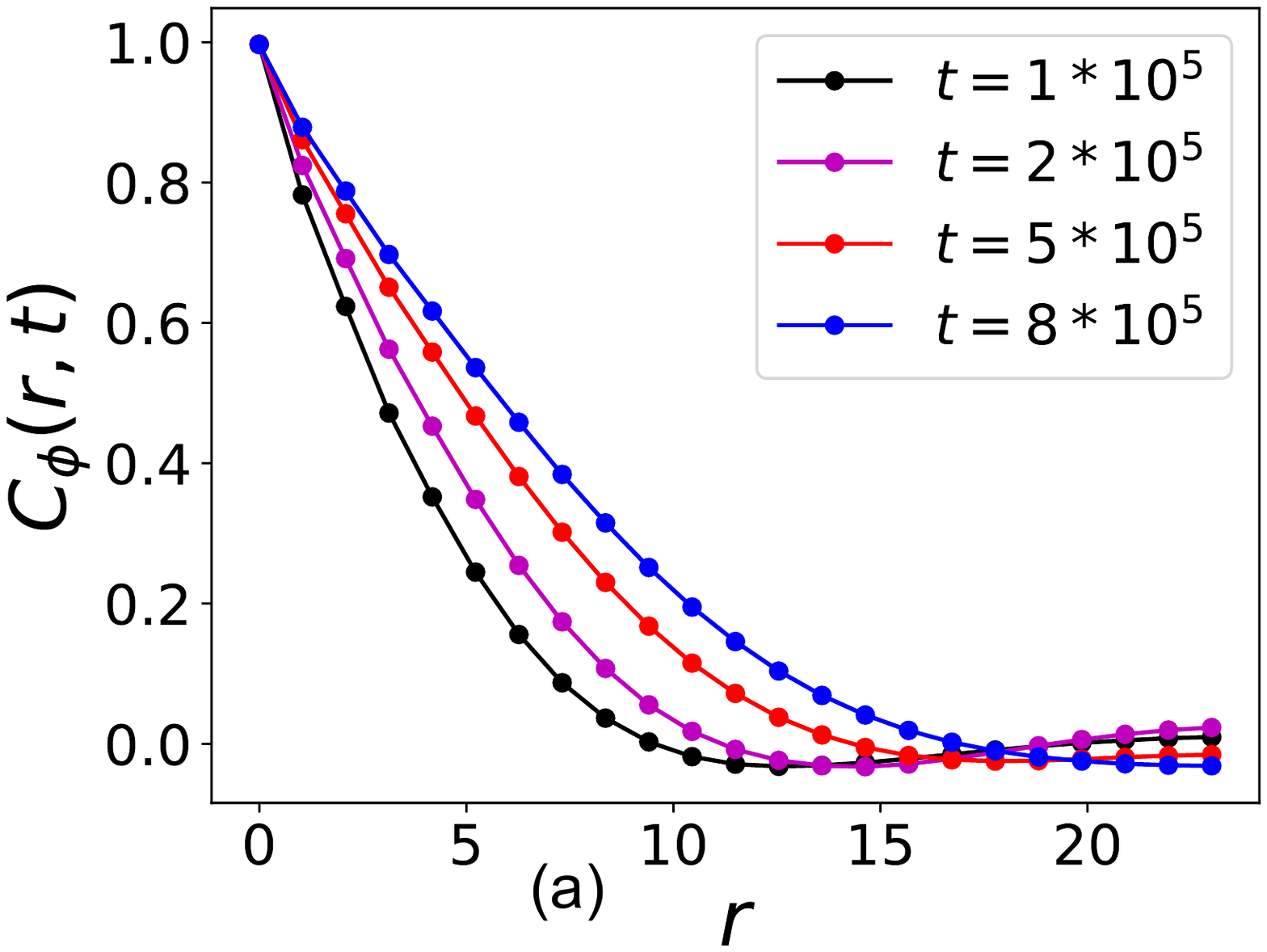}
}
\subfigure[]{

  \includegraphics[width=0.23\textwidth]{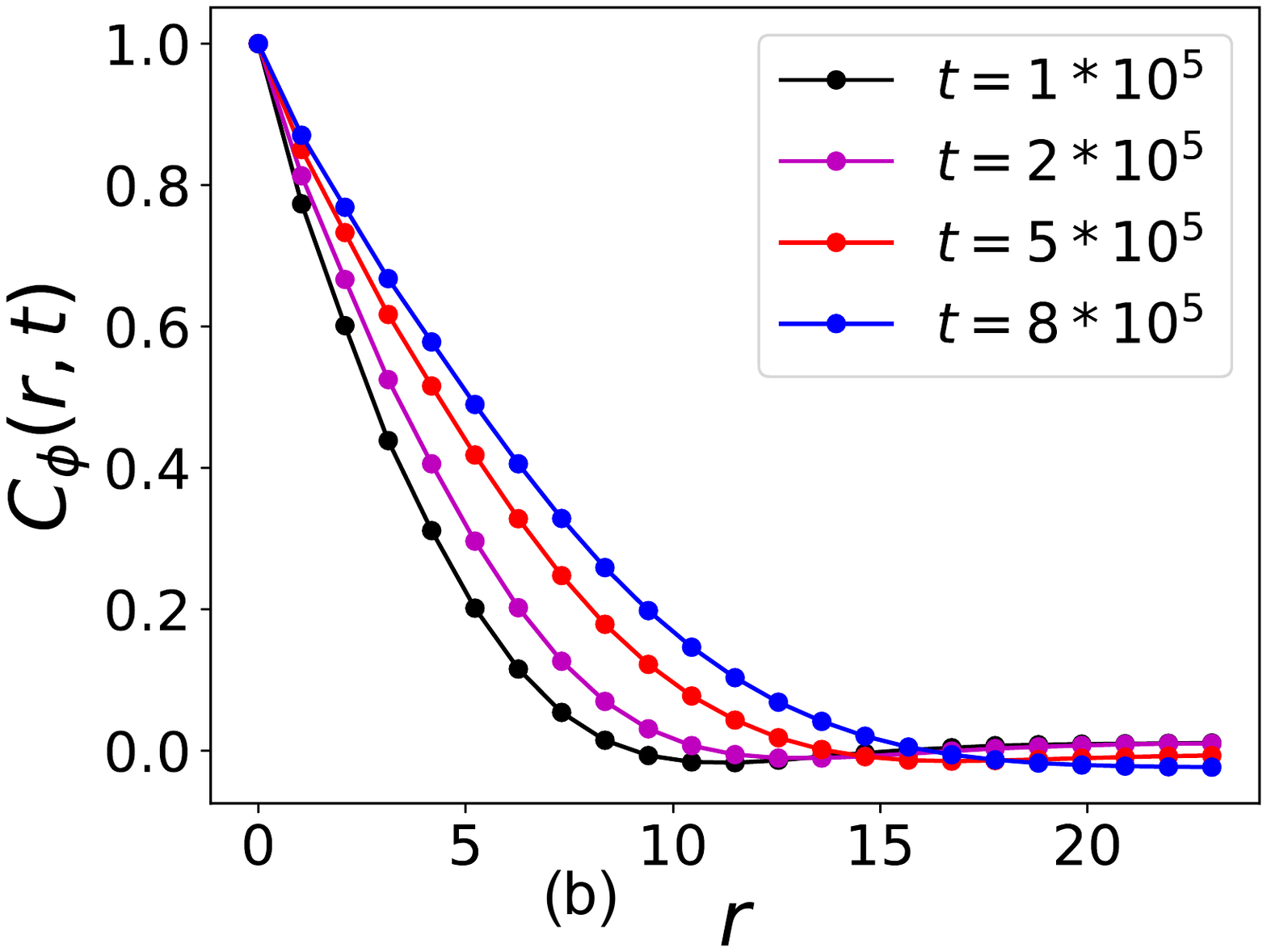}

}
\subfigure[]{

  \includegraphics[width=0.23\textwidth]{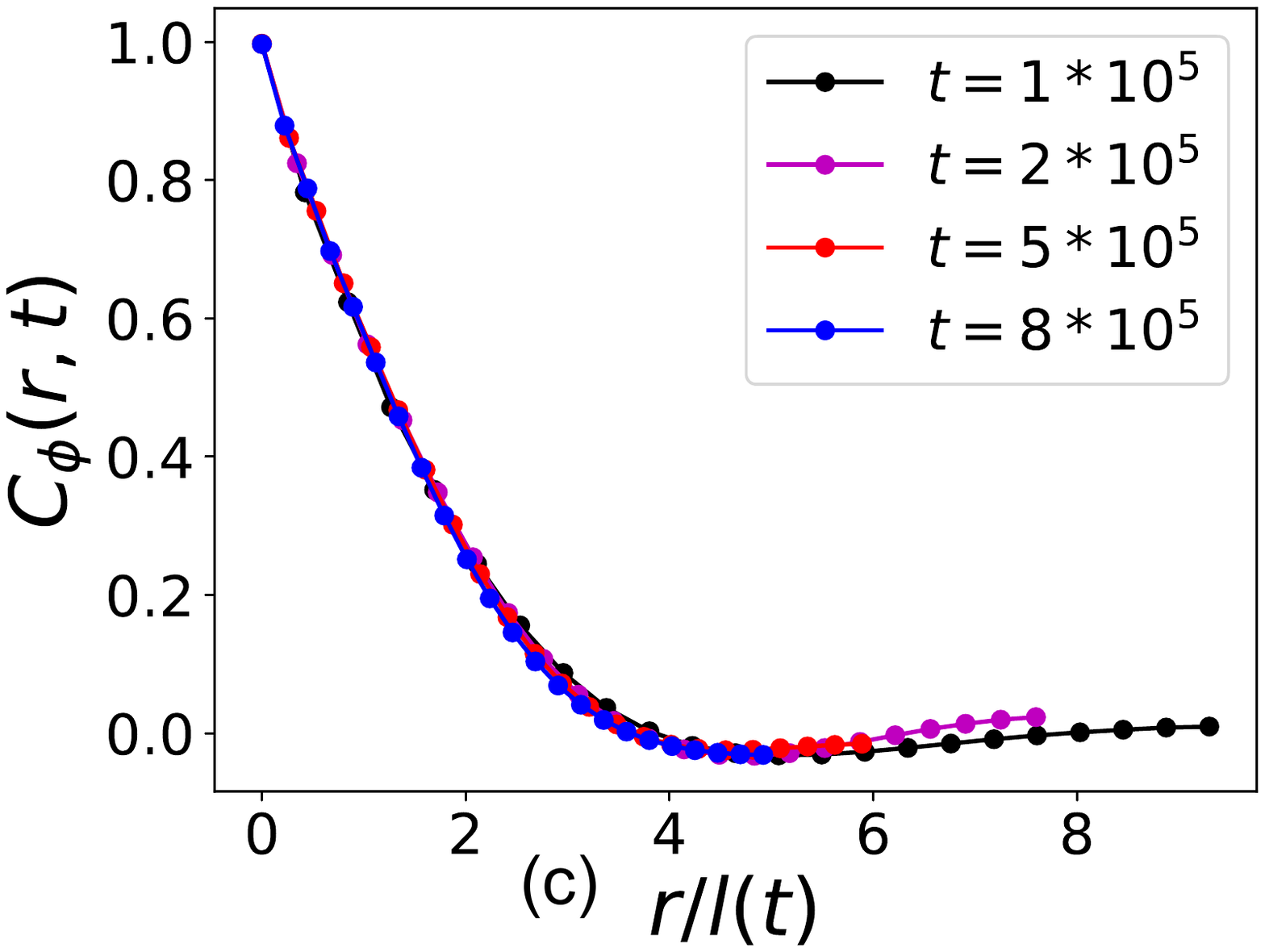}

}
\subfigure[]{

  \includegraphics[width=0.23\textwidth]{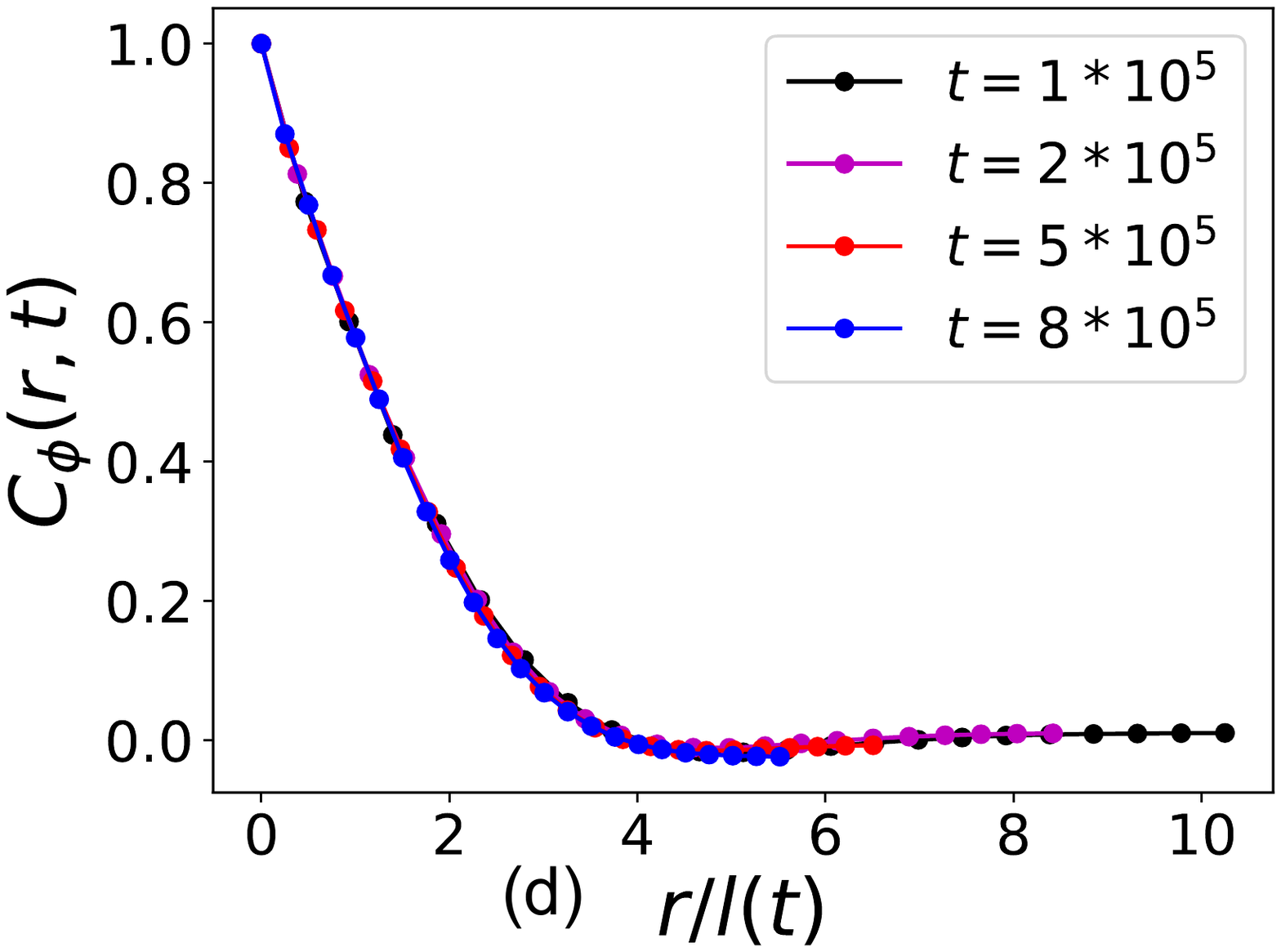}

}
\caption{(a) and (b) Plots of two-point spatial correlation function of \(\phi\), \(C_{\phi}(r,t)\) as function of distance between the points \(r\) at density \(\rho^*=0.8\) for quench temperature \(T_h^*=25\) and  \(T_h^*=40\) respectively, at different instants of time in 3D. The correlation function \(C_{\phi}(r,t)\) decays slowly as time progresses. (c) and (d) Plots of two-point spatial correlation function \(C_{\phi}(r,t)\) as function of distance \(r\) rescaled by the characteristic length \(l(t)\) at density \(\rho^*=0.8\) for quench temperature \(T_h^*=25\) and  \(T_h^*=40\) respectively, at different instants of time. The correlation functions at different times overlap onto a master curve implying the self-similar nature of domain growth.}
\label{fig:9}
\end{figure*}

 \section{Model}\label{sec:model}
\subsection{Molecular Dynamics(MD) Simulations}\label{sec:md_sim}
 We examine a binary mixture hot and cold particles undergoing 2-TIPS, using MD simulations in 3D as well as 2D.
 We begin with a system of N=80,000 particles, comprising an equal mixture of hot and cold particles in both 3D and 2D periodic boxes. The particles interact with each other by Lennard-Jones(LJ) potential given by equation \ref{eqn:LJ}.
 \begin{equation}
    U_{LJ}(r)=4\epsilon\bigg[\bigg(\frac{\sigma}{r}\bigg)^{12}-\bigg(\frac{\sigma}{r}\bigg)^6\bigg]
\label{eqn:LJ}    
\end{equation}
Here, \(\epsilon\) is the strength of the interaction, \(\sigma\) is the diameter of the LJ particle, and \(r\) is the distance between the interacting particles. The results from MD simulations are presented in reduced units, where we take \(\sigma\), \(\epsilon\), and mass \(m\) of the particles as units of length, energy, and mass respectively. Hence, the density and temperature in reduced units are given by \(\rho^*=\rho \sigma^3\) and \(T^*=k_BT/\epsilon\), respectively.
The simulations were performed in NVT ensemble using LAMMPS \cite{LAMMPS} software. To introduce the Two-temperature scalar model, we assign half of the particles in the system to the cold thermostat and the remaining half to the hot thermostat. We use Nos\'e-Hoover thermostat \cite{evans1985nose} to maintain the temperature of both hot (\(T_h^*\)) and cold (\(T_c^*\)) particles. The time step of integration is \(\Delta t^*=0.0005\) and damping factor of thermostat is chosen as \(\tau_T=50*\Delta t\) as used in our previous works \cite{chari2019scalar,D3SM00796K}. All the simulations in the present work are carried out at a high density of \(\rho^*=0.8\).\\
In our previous study we found that due to the energy transfer from hot to cold particles via collisions, the measured temperature of cold particles (\(T_c^{eff*}\)) is higher than its thermostat value (\(T_c^*\)), while the opposite is true for hot particles. Using the effective temperatures, the activity \(\chi\) of the non-equilibrium system is defined as the relative effective temperature difference between hot and cold particles, \( \chi = \frac{T_h^{eff*} - T_c^{eff*}}{T_c^{eff*}} \), where $T_h^{eff*}$ is the measured temperature of the hot particles. The simulations \cite{chari2019scalar,D3SM00796K} revealed that hot and cold LJ particles undergo phase separation when the activity \(\chi\) exceeds a density-dependent critical activity \(\chi_{crit}\).\\
The phase diagram for 2-TIPS of LJ particles (first reported in ref.\cite{chari2019scalar}) in 3D in the activity \(\chi\) - density \(\rho\) phase space is depicted in FIG.\ref{fig:op_ss}(c). A brief overview of previous studies is given in appendix \ref{app:premd}.\\
To study the kinetics of 2-TIPS, we initially prepare the binary mixture in mixed or homogeneous state by assigning the same temperature \(T^*=2\) for both hot (\(T_h^*\)) and cold (\(T_c^*\)) particles and allowing the system to reach equilibrium for 4 million (4M) time steps. Then, we quench the binary mixture in the mixed state to the phase-separated state in the phase diagram (FIG.\ref{fig:op_ss}(c)) by instantaneously increasing the temperature of hot particles \(T_h^*\)  to the desired value. The black triangle in the phase diagram is the state point with \(T_h^*=2\), \(\rho^*=0.8\) and \(\chi=0\)  corresponding to the initial mixed state. We perform two quenches in the phase-separated region by instantaneously increasing the temperature of the hot particles to \(T_h^*=25\) (\(\chi=8.79\)) and \(T_h^*=40\) (\(\chi=13.39\)). The red and blue triangles in FIG.\ref{fig:op_ss} (c) correspond to these final quenched states. After the quench, we run the simulation for 1M time steps, during which we study the temporal evolution of the binary mixture from a homogeneous state toward the phase-separated steady state. We ran ten independent simulations for each quench, and the resulting dynamical quantities were averaged across these runs to ensure robust statistical estimates.  \\  
We also developed a hydrodynamic coarse grained model using two coupled model B equations for the slow variable of hot and cold particles.  
Here, the slow variables are the density fields developed of the hot and cold particles, coupled through interspecies interactions. In binary collisions, energy transfer from hot particles to cold particles boosts the motion of cold particles while slowing down hot particles, leading to phase separation between the two species. The coupling terms are designed to modify the effective temperatures of both the particle types. 

\begin{figure*}[hbt]
\centering
\subfigure[]{
  \includegraphics[width=0.31\textwidth]{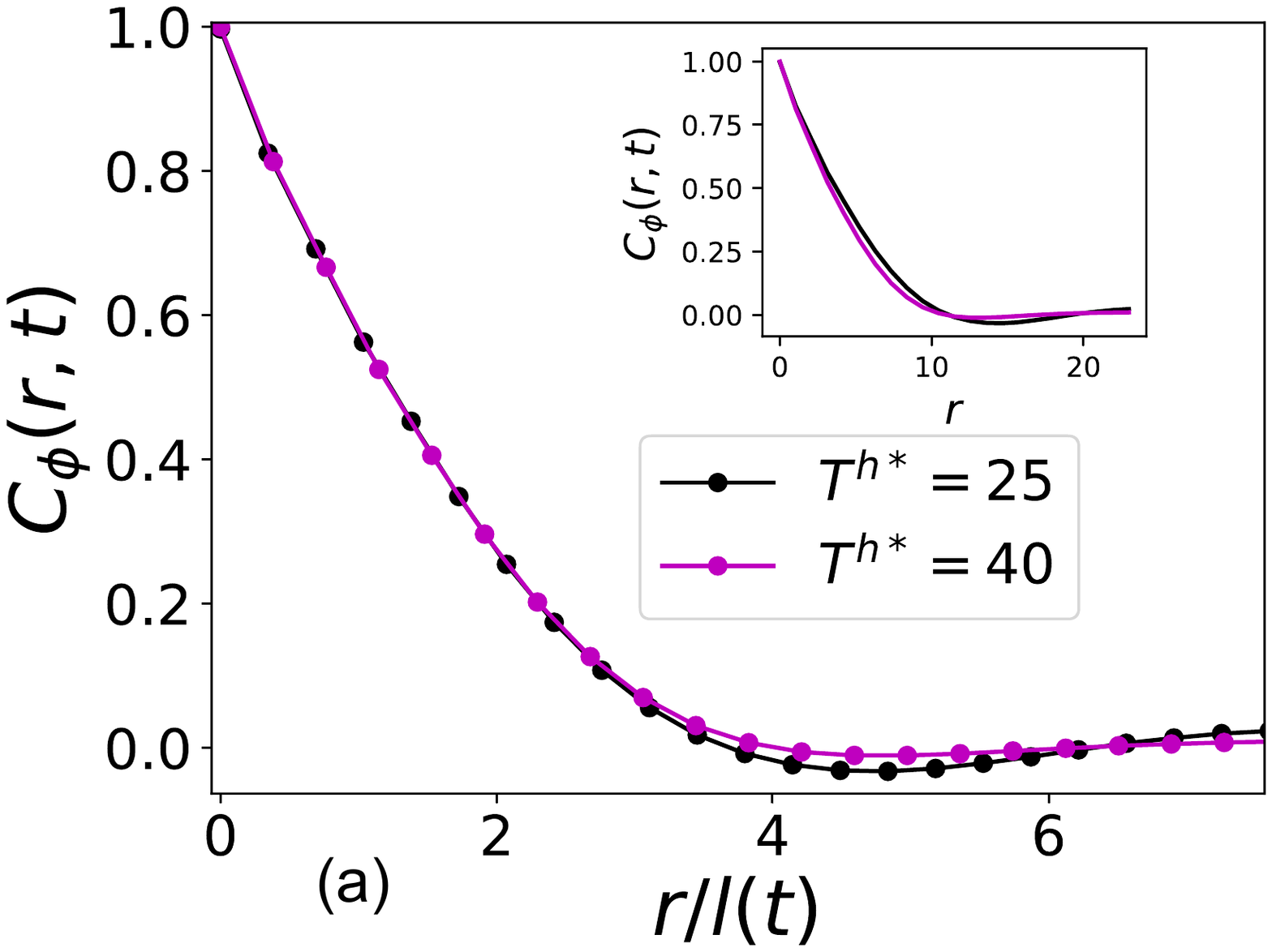}

}
~
\subfigure[]{
  \includegraphics[width=0.31\textwidth]{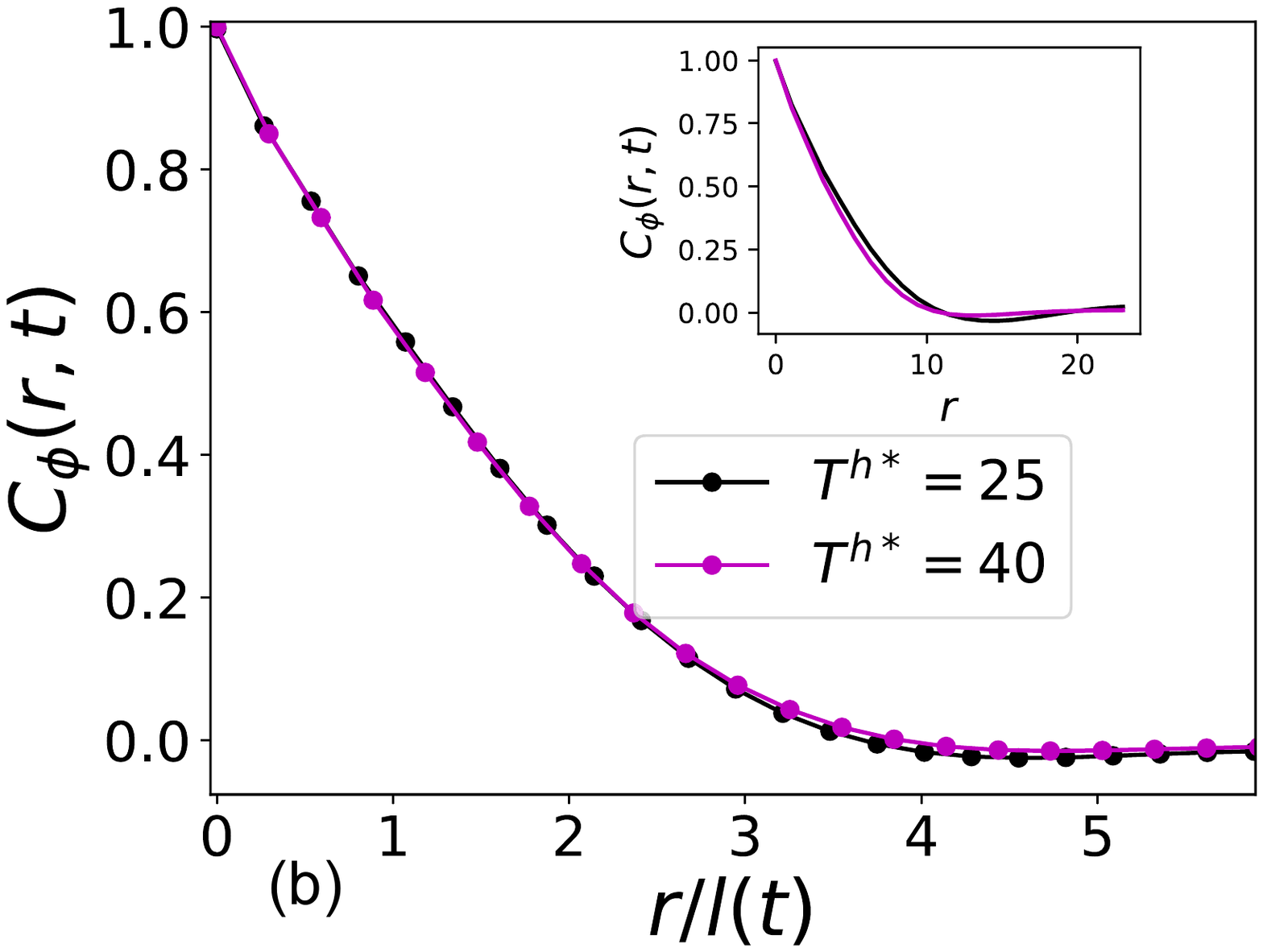}

}
\subfigure[]{
  \includegraphics[width=0.31\textwidth]{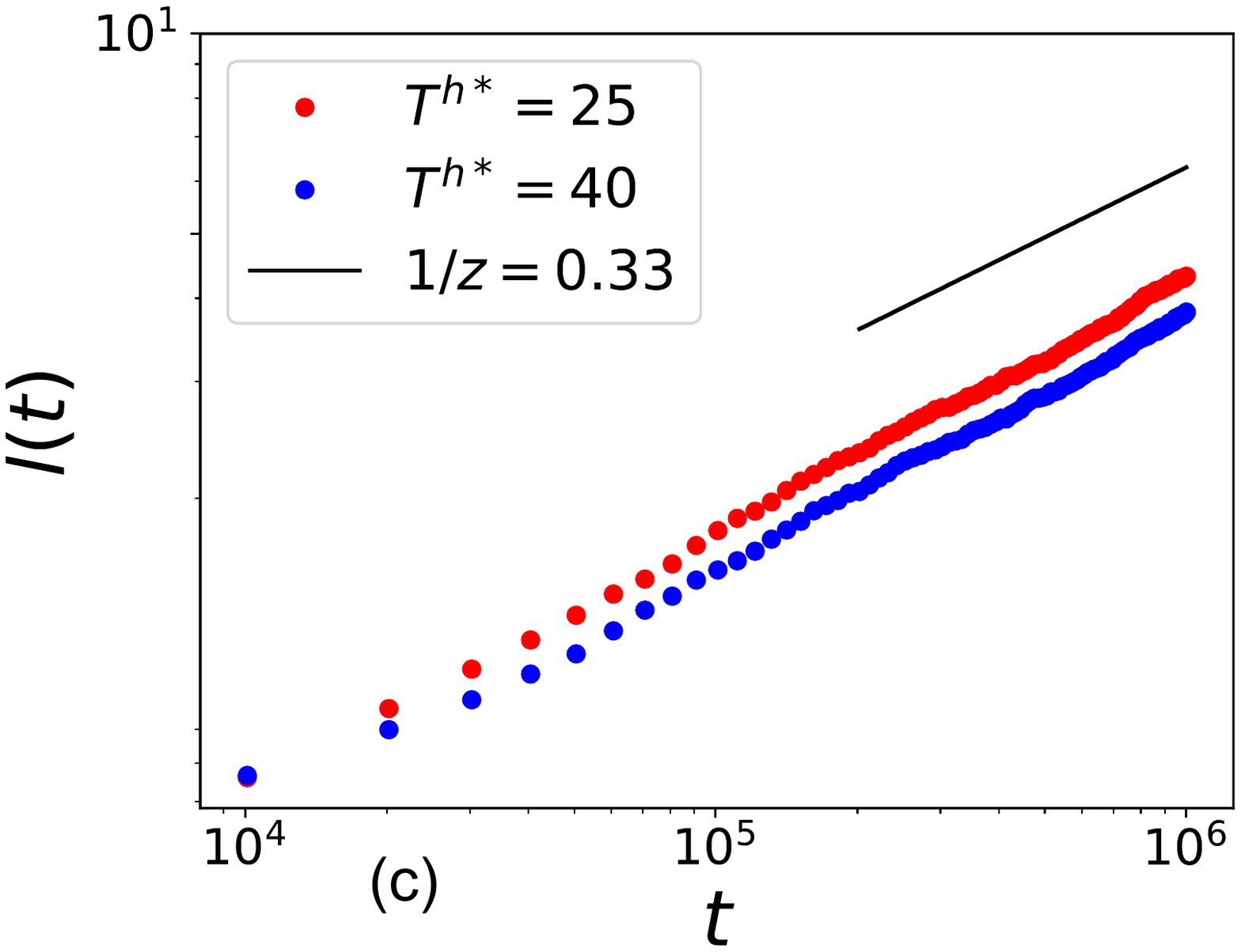}

}

\caption{(a) and (b) Plots illustrate static characteristic of \(C_{\phi}(r,t)\) as a function of rescaled distance \(r/l(t)\)  for different quench temperature at time \(t=2*10^5\) and \(t=5*10^5\), respectively, in 3D. The insets show \(C_{\phi}(r,t)\) as a function of distance \(r\) only. The plots reveal that domain structures at a given time for different quench temperatures are statistically similar.  (c) Log-Log plot of characteristic length \(l(t)\) versus time \(t\) shows algebraic growth of \(l(t)\). We see that the growth exponent \(1/z\) has a value close to 1/3 at late times.}
\label{fig:10}
\end{figure*}

\begin{figure*}[hbtp!]
\centering
  \includegraphics[width=0.41\textwidth]{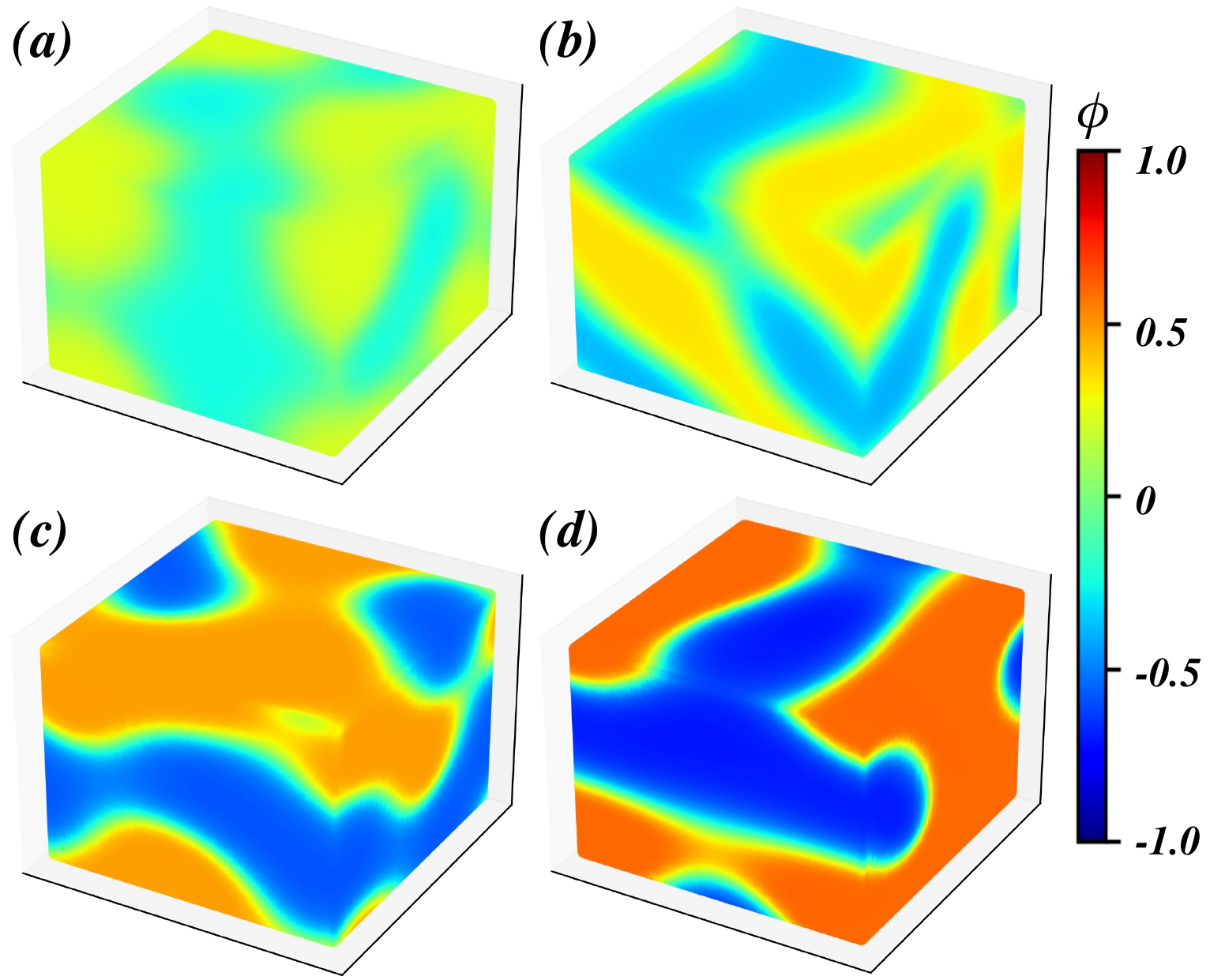}
  \includegraphics[width=0.55\textwidth]{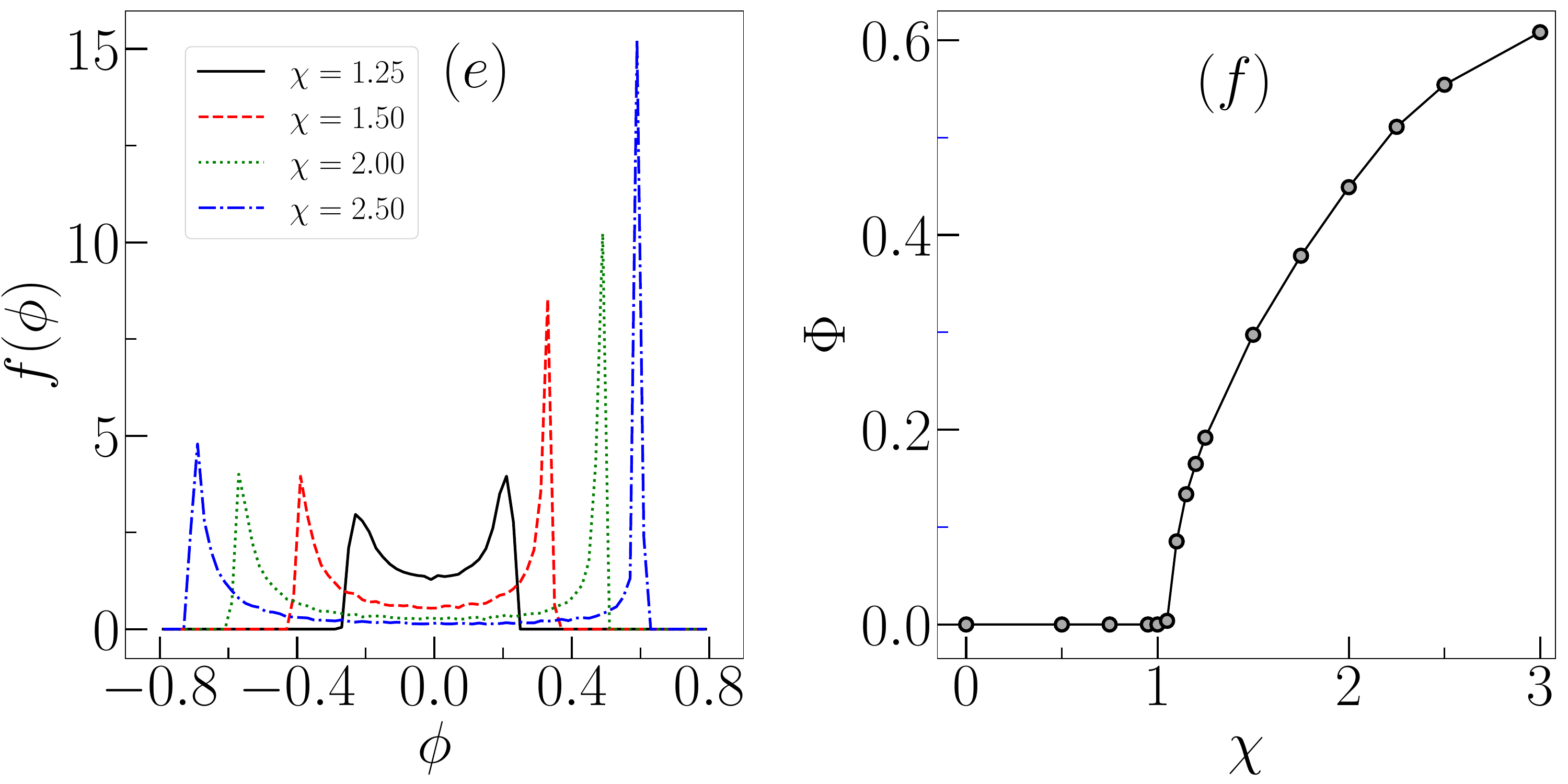}
\caption{The figure depicts the phase separation in critical mixture at different activity in 3D. Subplots (a)-(d) show the snapshot of $\phi$ in the steady state of the system for different values of activity: (a) $\chi = 1.25$, (b) $\chi = 1.50$, (c) $\chi = 2.00$, and (d) $\chi = 2.50$, respectively. The subplot (e) shows the probability distribution of $\phi$, $f(\phi)$ in the steady state for different values of activity. The subplot (f) shows the variation of PSOP, $\Phi$, with activity, $\chi$, in the critical mixture. Parameters : System size, $L = 32 $ for (a)-(d) , and $L = 64$ for (e) and (f) }
\label{fig:1}
\end{figure*}

\subsection{Coarse Grained (CG) model}\label{sec:cg}
In a system where the total number of particles is conserved, the density field $\rho(\boldsymbol{r},t)$ acts as a slow variable \cite{forster2018hydrodynamic}. The conserved order parameter field corresponding to $\rho(\boldsymbol{r},t)$ is denoted by $\psi(\boldsymbol{r},t)$ such that $\psi(\boldsymbol{r},t)>0$ ($\psi(\boldsymbol{r},t)<0$) implies a region enriched (depleted) in particles and vice versa. According to the Halprin-Hohenberg classification \cite{hohenberg1977theory}, the dynamics of a system with conserved order parameter is described by Model B. The free energy functional, $\mathcal{F}[\psi(\boldsymbol{r},t)]$, for a $D-$dimensional system is given by,
\begin{equation*}
    \mathcal{F}[\psi(\boldsymbol{r},t)] = \mathlarger{\int} \bigg{\{}\frac{\alpha}{2} \psi^2  + \frac{\beta}{4} \psi^4 + \frac{\gamma}{2} (\nabla\psi)^2\bigg{\}} d^Dr
\end{equation*}
where, $\alpha = \alpha_0 (\frac{T - T_o}{T_o})$ with $T_o$ being the critical temperature. For $\alpha > 0$ the homogeneous state is stable while for $\alpha < 0$ the homogeneous state is unstable. Both $\beta$ and $\gamma$ are positive constants, ensuring the stability of the model.\\
The dynamics of $\psi(\boldsymbol{r},t)$ is governed by the continuity equation,  
\begin{align}
     &\quad \frac{\partial \psi(\boldsymbol{r},t)}{\partial t} = -\Vec{\nabla} \cdot \Vec{J} \\
     &\quad \Vec{J} = -\Vec{\nabla}\mu \\
     &\quad \mu = \frac{\delta \mathcal{F}}{\delta \psi} \\
     &\quad \frac{\partial \psi(\boldsymbol{r},t)}{\partial t} = \nabla^2\left(\frac{\delta \mathcal{F}}{\delta \psi}\right)
\end{align}
where, $\Vec{J}$ is the current and $\mu$ is the chemical potential. We have two sets of Model B equations, one for each of the hot and cold particles, and these two sets are coupled. The order parameter for hot and cold are denoted by $\psi_c$ and $\psi_h$, respectively. As already discussed in Sec.\ref{sec:md_sim}, as a result of the interaction between the hot and the cold particles, the effective temperature of both type of particles ($T_h^{eff*}$ and $T_c^{eff*}$, 
 respectively) are different from their original temperatures : $T_h^*>T_h^{eff*}>T_c^{eff*}>T_c^*$ \cite{grosberg2015nonequilibrium,grosberg2018dissipation,chari2019scalar}. Following these results, we introduce the coupling between the hot and cold particles in such a manner that it reduces the effective temperature of hot (i.e. $T_h^*>T_h^{eff*}$) and raises the effective temperature of cold (i.e. $T_c^*<T_c^{eff*}$). The free energy functional for the hot ($\mathcal{F}_h$) and cold ($\mathcal{F}_c$) subsystems are as follows
\begin{equation}
    \mathcal{F}_c[\psi_{c,h}(\boldsymbol{r},t)] = \mathlarger{\int} \bigg{\{}\frac{\alpha_c}{2} \psi_c^2  + \frac{1}{4} \psi_c^4 + \frac{1}{2} (\nabla\psi_c)^2 +\frac{1}{2}\psi_c^2 \psi_h \bigg{\}} d^Dr
    \label{eq:1}
\end{equation}
and 
\begin{equation}
    \mathcal{F}_h[\psi_{c,h}(\boldsymbol{r},t)] = \mathlarger{\int} \bigg{\{}\frac{\alpha_h}{2} \psi_h^2  + \frac{1}{4} \psi_h^4 + \frac{1}{2} (\nabla\psi_h)^2 -\frac{1}{2}\psi_h^2 \psi_c \bigg{\}} d^Dr
    \label{eq:2}
\end{equation}

In the above expressions, $\alpha_c = \bigg{(}\frac{T_c^* - T_{o,c}^*}{T_{o,c}^*}\bigg{)}$ and $\alpha_h = \bigg{(}\frac{T_h^* - T_{o,h}^*}{T_{o,h}^*}\bigg{)}$, where $T_{o,c}^*$ and $T_{o,h}^*$ are the critical temperature for cold and hot particles, respectively, and all other coefficients are set to $1$ in reduced units. In Eq.\ref{eq:1} and Eq.\ref{eq:2}, the last term describe the coupling of hot into cold and vice versa. These couplings control the energy transfer between a hot and a cold particle during their collision, ignoring the other means of energy loss during the collision.\\
The temperatures of hot and cold are chosen such that the temparature of cold is below the critical temparature $T_c^* < T_{o,c}^*$ (i.e. $\alpha_c<0$), while the temperature of hot is above the critical temperature $T_h^* > T_{o,h}^*$ (i.e. $\alpha_h>0$). The form and sign of coupling is chosen such that it directly modifies the temperature of the hot and the cold species, which is the leading order effect.   However, sub-leading order effects can be incorporated by choosing the form of coupling such that it modifies the surface tension term in the free energy. But for a system undergoing the macroscopic phase separation, the coupling which modifies the surface tension term is less relevant for large length and times scales. The different signs of couplings in hot and cold particles make the system intrinsically non-equilibrium in nature.  \\
The combined free energy of the mixed system is given by, $\mathcal{F}_{mix} = \mathcal{F}_h + \mathcal{F}_c$. The dynamical equations for the evolution of $\psi_c(r,t)$ and $\psi_h(r,t)$ are as follows
\begin{widetext}
\begin{equation}
    \frac{\partial \psi_c(\boldsymbol{r},t)}{\partial t} = \nabla^2 \bigg{(}\frac{\delta \mathcal{F}_{mix}}{\delta \psi_c}\bigg{)} \\ \Rightarrow \frac{\partial \psi_c(\boldsymbol{r},t)}{\partial t} = \nabla^2 \bigg{[}  \alpha_c\psi_c +  \psi_c^3 -  \nabla^2 \psi_c + \psi_h \psi_c - \frac{1}{2}\psi_h^2 \bigg{]} 
    \label{eq:3}
\end{equation}

\begin{equation}
    \frac{\partial \psi_h(\boldsymbol{r},t)}{\partial t} = \nabla^2 \bigg{(}\frac{\delta \mathcal{F}_{mix}}{\delta \psi_h}\bigg{)} \Rightarrow \frac{\partial \psi_h(\boldsymbol{r},t)}{\partial t} = \nabla^2 \bigg{[} \alpha_h\psi_h +  \psi_h^3 -  \nabla^2 \psi_h -  \psi_c \psi_h + \frac{1}{2} \psi_c^2 \bigg{]}
    \label{eq:4}
\end{equation}
\end{widetext}
For simplicity, we fixed the critical temperatures of both species $T_{o,c}^* = T_{o,h}^* = T_o^* = 0.50$. The control parameter is defined as $\chi_{cg} = \alpha_h - \alpha_c$, which is a measure of the relative temperature difference between the two species and is equivalent to the $\chi$ introduced in MD simulation details. Later we use the same symbol $\chi$ for the activity parameter in MD and CG models. We performed the numerical integration of  Eq.\ref{eq:3} and Eq.\ref{eq:4} with periodic boundary conditions in all directions. For $D=2$, we simulate the system on a square lattice of size $L \times L$, and for $D = 3$  we simulate the equations on a cubic lattice of size $L \times L \times L$. For the numerical integration, we used the Euler integration scheme with space and time grid size $\Delta x = 0.5$ and $\Delta t = 0.01$, respectively.
We used small step sizes, performed initial checks, and maintained the stability criterion $\frac{(\Delta x)^2}{\Delta t} < \frac{1}{2}$. In the Model B language, the system is referred to as critical or off-critical depending on the value of $\psi_o = <\psi(\boldsymbol{r},0)>_{\boldsymbol{r}}$, where $<...>_{\boldsymbol{r}}$ signifies average over all the lattice points: $\psi_o = 0$ for critical and $\psi_0 \neq 0$ for off-critical \cite{bray1994theory,puri2009kinetics} composition of the system.  In the context of our coarse grained model for 2-TIPS, the critical and off-critical mixtures are defined as ($\psi_{o,c} = 0$ and $\psi_{o,h} = 0$) and   ($\psi_{o,c} \ne 0$ and $\psi_{o,h} \ne 0$) respectively.  The initial condition consists of $\psi_{c/h}(r,t)$ having small fluctuation around $\psi_{o,c/h}$. 
At $t=0$, the cold and hot systems are instantaneously quenched to temperatures $T_c^* (< T_o^*)$ and $T_h^* (> T_o^*)$, respectively, and are allowed to evolve. A single simulation step consists of updating $\psi_c$ and $\psi_h$  at every lattice point. The results shown here are obtained for a square box of size $L = 256$ and $512$ for 2D and a cube of size $L = 32$ and $64$ for 3D. For 2D system, a total of $5 \times 10^6$ time steps are considered. The quantities describing the steady state of the system are calculated after $3 \times 10^6$ time steps by which time the system is observed to have reached it's steady state. Similarly, for the 3D system, a total of $6 \times 10^5$ simulation steps are considered, and the observables describing the steady state are calculated after $3 \times 10^5$ time steps. For better statistical accuracy, we average over $200$ independent realizations for 2D and $150$ independent realizations for 3D.\\ 
In this article, we show the results for the critical mixture, which compares well with the high density limit of the microscopic model. The results for the off-critical mixture will be discussed in future studies.  \\

To quantify the phase separation between the cold and the hot particles, we define an order parameter $\phi(\boldsymbol{r}) = \psi_h(\boldsymbol{r}) - \psi_c(\boldsymbol{r})$, which contains the information of phase separation on a local scale. A measure of global phase separation in the system can be obtained by calculating, $\Phi = \frac{1}{2L^D}\mathlarger{\sum}_{\boldsymbol{r}} \vert \phi(\boldsymbol{r}) \vert$,  the Phase Separation Order Parameter (PSOP)  as defined in MD \cite{chari2019scalar,C9SM02552A} (appendix \ref{app:premd}). \\

\begin{figure*}[hbtp!]
\centering
 \includegraphics[width=0.78\textwidth]{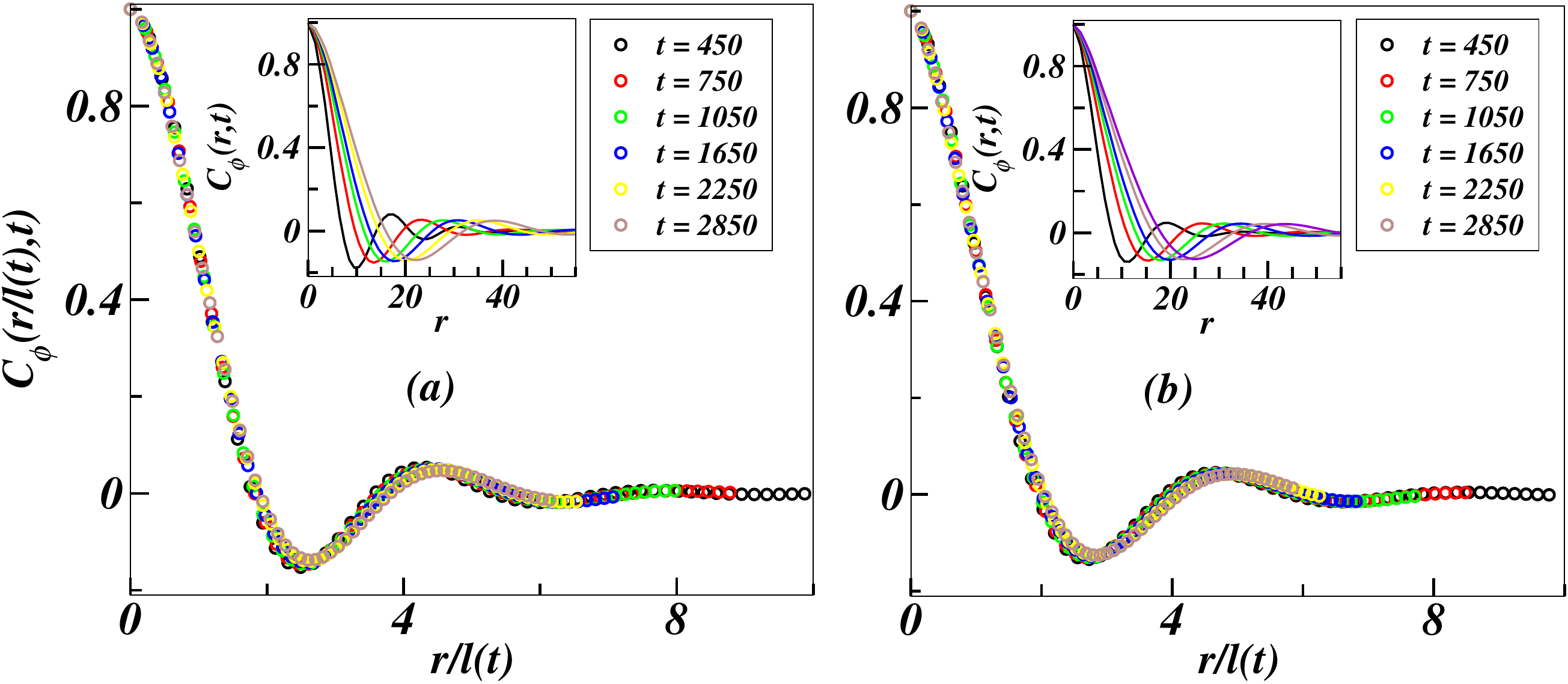}
 \includegraphics[width=0.78\textwidth]{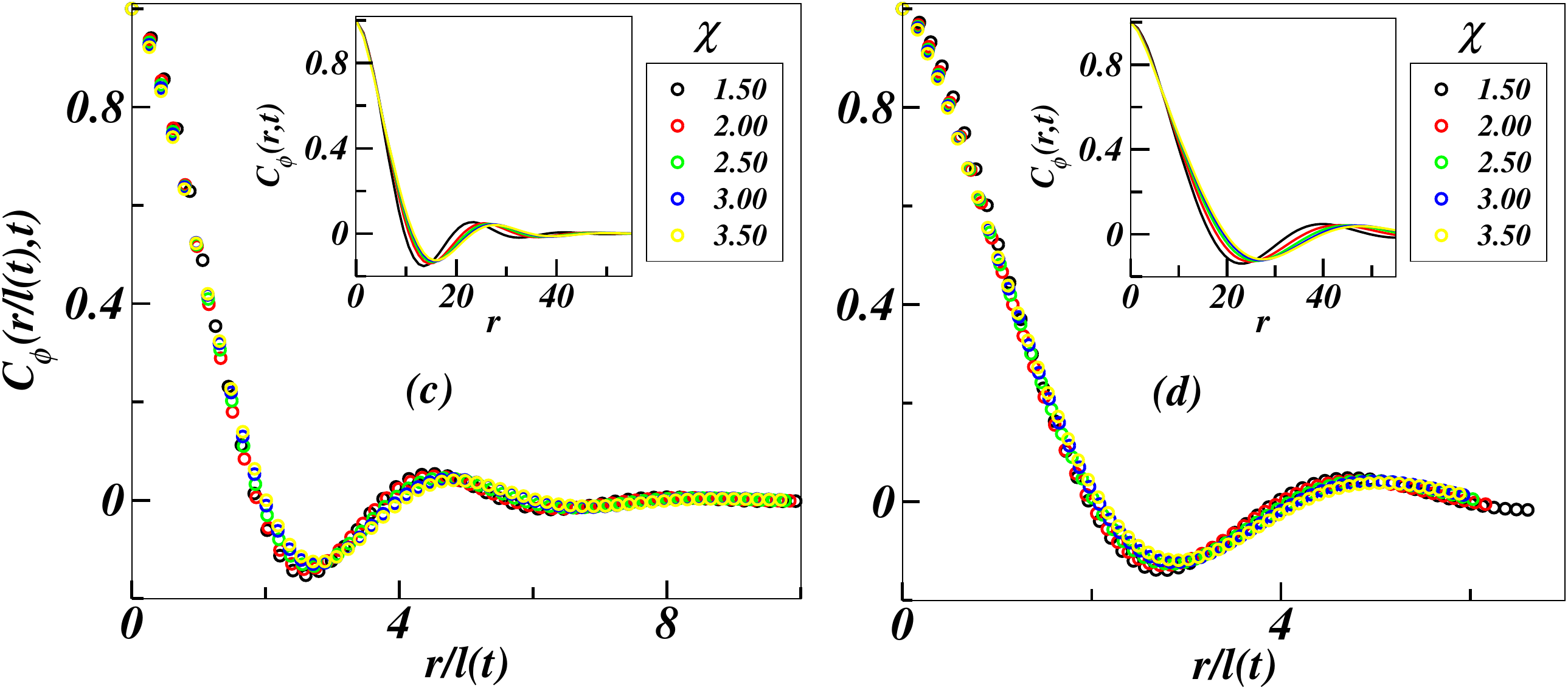}
\caption{The figure illustrates the behavior of the two point correlation function of $\phi$, $C_{\phi}(r,t)$, for the mixture with critical composition, $\psi_{0,c/h} = 0$, in 3D. In all the subplots (a-d), the color code is the same for the main plot and inset. Subplots (a) $\&$ (b) show the dynamic characteristics of $C_{\phi}(r,t)$ at different times for two different activity : (a) $\chi = 1.50$, (b) $\chi = 2.50$. The insets show $C_{\phi}(r,t)$ at different times and in the main plot window we show the scaling behavior of correlation function when the distance, $r$, is rescaled by the characteristics length scale, $l(t)$. The subplots (c) $\&$ (d) show the static characteristics of $C_{\phi}(r,t)$ for different activity at two time instants (c) $t = 450$, (d) $t = 2550$. The insets show the $C_{\phi}(r,t)$ for different activity and the main plots show the scaling behavior of the correlation function when plotted against the rescaled distance $r/l(t)$. System size, $L = 128$.}
\label{fig:3dscaling}
\end{figure*}

\begin{figure*}[hbtp!]
\centering
 \includegraphics[width=0.78\textwidth]{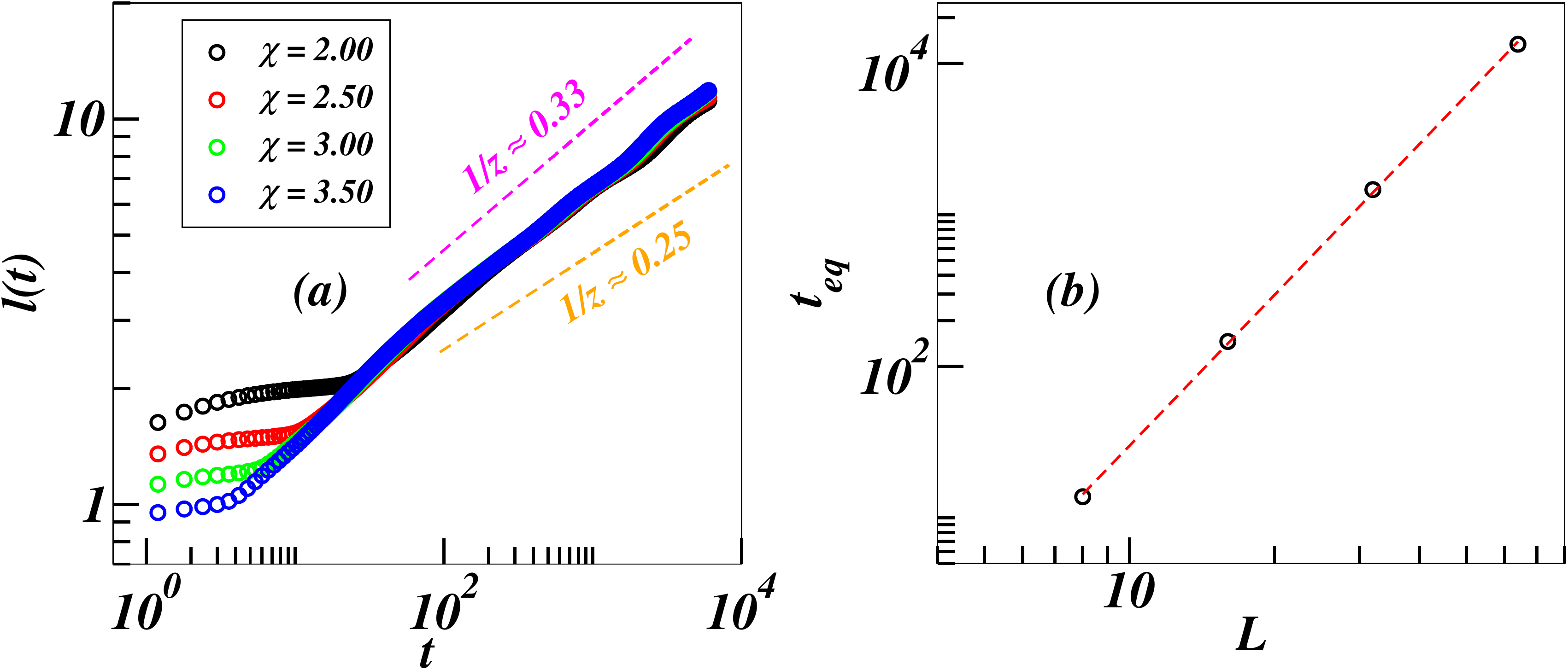}
 \includegraphics[width=0.78\textwidth]{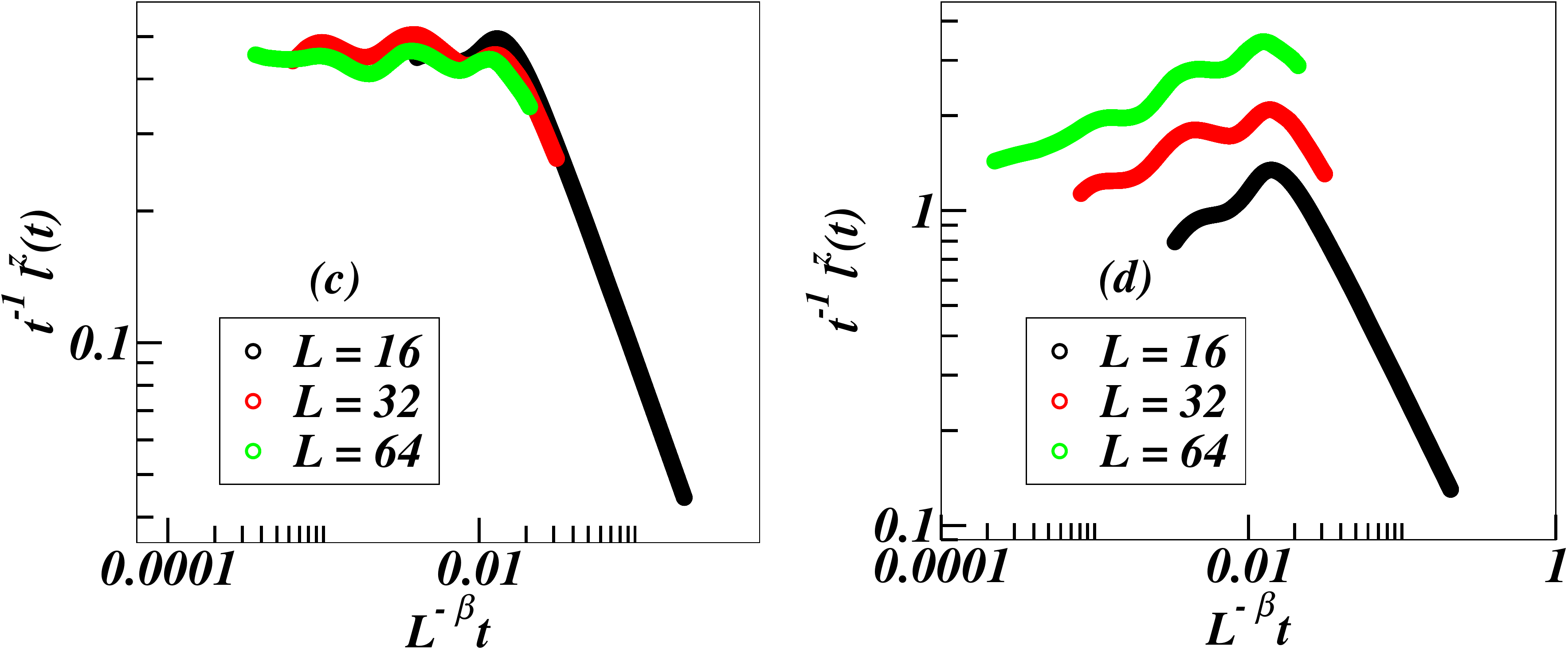}
\caption{The figure depicts (a)Log-Log plot of the correlation length, $l(t)$ {\em vs.} $t$ at different activity for the mixture with critical composition, $\psi_{0,c/h} = 0$, in 3D. The dashed lines labeled  $1/z = 0.33$ and $0.25$ denotes growth law $l(t) \sim t^{1/3}$ and $l(t) \sim t^{1/4}$, respectively. System size, $L = 64$; (b) Log-Log plot of $t_{eq}$ $vs.$ system size, $L$, for activity $\chi = 2.50$. The red dashed line shows the powerlaw fit $t_{eq} \sim L^{\beta}$ with exponent $\beta = 3.3092$; (c)-(d) Log-Log plot of $t^{-1}l^{z}(t,L)$ $vs.$ $L^{-\beta} t$ for different system sizes for $\chi = 2.50$ for two different values of $1/z$: (c) $1/z \approx 0.32$, and (d) $1/z = 0.25$. The value of $\beta$ is taken from the subplot (b).}
\label{fig:3}
\end{figure*}
\section{Results}\label{sec:res}

\subsection{2-TIPS in 3-dimensions}\label{sec:md_3d}
\underline{\textit{\text{Kinetics of phase separation}}}: 
The phase separation of the binary mixture starts immediately after quench and proceeds by the formation and growth of bi-continuous domains, which are rich in either hot or cold particles, reminiscent of spinodal decomposition in purely passive systems. The series of snapshots in FIG.\ref{fig:con3d} shows the instantaneous configurations of the evolution of 80,000 hot and cold particles at various instances of time after the hot particles are quenched to \(T_h^*=25\) at density \(\rho^*=0.8\). The initial step towards determining the size of the phase separating domains is to quantify the domain morphologies by an order parameter field \(\phi(r,t)\), which can clearly distinguish the hot and cold domains. The cold domains exhibit a density higher than the average density, while hot domains have a lower density than the average density, a feature consistently exhibited by various systems undergoing 2-TIPS. So, the phase separating hot and cold domains not only exhibit differences in concentrations of hot and cold particles but also in their densities, either of which could be used to differentiate them. However, as explained in appendix \ref{app:premd}, due to the trapping of hot particles in the cold domains, especially in 2D, using the number difference between the hot and cold particles (\(n_h-n_c\)) as the order parameter underestimates the domain size. So, the variation in densities of hot and cold domains is used as a criterion to distinguish them. From the MD trajectories, the order parameter field \(\phi(r,t)\) is obtained as follows: We discretize the position of particles in the simulation volume by mapping them to a lattice with spacing \(\sigma\), which is equal to the diameter of LJ particles. Then, we calculate the local density \(\rho(r)\) at each point by counting the number of nearest neighbors. We obtain the order parameter field by assigning \(\phi=1\) to the points whose local density \(\rho(r)>0.8\) (dense cold regions) and \(\phi=-1\) to the points whose local density \(\rho(r) \le 0.8\) (dilute hot regions). FIG.\ref{fig:8} illustrates the order parameter field \(\phi(r,t)\) for quench temperature \(T_h^*=25\) and density \(\rho^*=0.8\) at various instants of time $t$. The order parameter field \(\phi(r,t)\) distinctly delineates the regions with the majority cold (\(\phi(r)=1\)) and hot particles (\(\phi(r)=-1\)). FIG.\ref{fig:8} demonstrates the growth in the bicontinuous hot and cold domains with the progression of time.   \\
To capture the length scale of the phase separating domains, we calculate the two-point spatial correlation function of order parameter \(\phi\), $C_{\phi}(r,t)$ defined as 
\begin{align}
    C_{\phi}(r,t) &= \langle \phi(\boldsymbol{r}_0,t)\phi(\boldsymbol{r}_0 + \boldsymbol{r},t) \rangle \notag \\
    &\quad - \langle \phi(\boldsymbol{r}_0,t) \rangle \langle \phi(\boldsymbol{r}_0 + \boldsymbol{r},t) \rangle
    \label{eq:cor}
\end{align}

where $<...>$ implies the averaging over reference point $\boldsymbol{r}_0$ and multiple independent realizations.
The correlation/characteristic length, denoted by $l(t)$, provides a measure of the average size of the domains at time $t$ and it can be calculated from the decay of $C_{\phi}(r,t)$ by setting a cut-off as $C_{\phi}(l(t),t) = a$. In this study, we use $a = 0.5$. However, the statistical and the scaling properties of $C_{\phi}(r,t)$ and $l(t)$ remain consistent regardless of the choice of $a$. FIG.\ref{fig:9}(a-b) shows the plots of two-point spatial correlation function of \(\phi\) versus distance between the points \(r\) at density \(\rho^*=0.8\), for quench temperatures \(T_h^*=25\) and  \(T_h^*=40\) respectively, at different timesteps $t$ after quench. We discern that the correlation function decays more slowly as time progresses, implying the increasing length of domains with time. When the distance \(r\) is rescaled by the characteristic length \(l(t)\), the correlation functions \(C_{\phi}(r,t)\) at different instants of time collapse onto a single master curve as depicted in FIG.\ref{fig:9}(c-d). This demonstrates that the domain morphologies of phase separating binary mixture at different times are statistically self-similar, solely distinct by their length scale.\\
We also plot correlation function \(C_{\phi}(r,t)\) at a given time for different quench temperatures in FIG.\ref{fig:10} to demonstrate the static characteristics of domain growth. FIG.\ref{fig:10}(a-b) illustrate \(C_{\phi}(r,t)\) as a function of rescaled distance \(r/l(t)\) for different quench temperatures, at time \(t=2\times10^5\) and \(t=5\times10^5\), respectively. Again, we observe that \(C_{\phi}(r,t)\) for different quench temperatures at a fixed time collapse onto a single master curve, indicating the statistical similarity between domain structures at different quench temperatures.\\ 
 Further, we compare the results of MD simulation with the CG model. FIG.\ref{fig:critsnaps} illustrates the system's evolution from an initially homogeneous state in CG model. In the early stages, hot and cold particles are well-mixed, causing $\phi$ to fluctuate slightly around zero across the system. As time progresses, phase separation sets in, with regions of $\phi > 0$ ($\phi < 0$) corresponding to regions rich in hot (cold) particles, respectively (The signs of \(\phi\) for hot and cold regions in CG model is opposite to the convention followed in MD results. However for the symmetric mixture as here  interchanging the signs does not affect our observation). During the intermediate time regime, bi-continuous domains emerge, reminiscent of the domain structure in the critical composition, while at later times, we observe complete phase separation between the hot and cold particles. Further, the dependence of the phase separation on the activity is shown in FIG.\ref{fig:1}. In FIG.\ref{fig:1}(a-d) we show the late time snapshot of $\phi$ for different values of activity. The snapshots clearly show that the phase separation is much stronger at larger activity values. We also calculated the probability density function (PDF) of $\phi$ as $f(\phi) = \frac{P(\phi)}{\Delta \phi}$, where $P(\phi)$ is the probability that the value of $\phi$ for a randomly selected lattice point lies with the interval $[\phi - \frac{\Delta \phi}{2}, \phi + \frac{\Delta \phi}{2})$. \\
 FIG.\ref{fig:1}(e) depicts the plot of $f(\phi)$, showcasing two distinct peaks that signify phase separation in the system. At low activity, these peaks are broad and closely spaced in the PDF, with the peak values of $\phi$ corresponding to the two phases. The broadening of the peaks may result from fluctuations of $\phi({\bf r})$ around its mean value in the bulk phases, as well as from the presence of a diffuse interface between the two phases ($\phi >0$ and $\phi<0$). As the activity increases, the peaks sharpen and the separation between them widens, indicating more pronounced phase separation between the cold and hot regions, along with the emergence of a sharper interface between the phases ($\phi >0$ and $\phi<0$). FIG.\ref{fig:1}(f) shows the variation of (PSOP), $\Phi$, with activity, $\chi$, exhibiting an increasing trend similar to that in MD studies \cite{chari2019scalar}.\\
 To characterize the growth of phase separating regions, we compute the two-point correlation function of $\phi({\bf r})$, as described by Eq.\ref{eq:cor}. The correlation function is averaged over $150$ independent realizations.  
The growth of domains over time is reflected in the slower decay of $C_{\phi}(r,t)$ as shown in the insets of FIG.\ref{fig:3dscaling}(a-b). The correlation length, $l(t)$, is calculated as the $0.5$ crossing of $C_{\phi}(r,t)$ as described for MD. The correlation function at different times collapse on to a single master curve when the distance, $r$, is rescaled by $l(t)$ as shown in the main frame in FIG.\ref{fig:3dscaling}(a-b). This suggests the existence of a single characteristic length scale in the system, given by the $l(t)$, and describes the statistical similarity of the domains at different times when $r$ is scaled by $l(t)$. Further, FIG.\ref{fig:3dscaling}(c-d) show the static characteristics of the domain growth process by comparing $C_{\phi}(r,t)$ for different activity, $\chi$, at a fixed time , when distance is scaled by $l(t)$. The insets of FIG.\ref{fig:3dscaling}(c-d) show the plot of $C_{\phi}(r,t)$, while the main frame shows the scaling collapse of the correlation function with respect to $\chi$. This suggests that the morphology of the domains at a fixed time is statistically similar for different activity, $\chi$, and they differ merely by the length scale. \\

\underline{\textit{Growth exponent}}: The length of the phase separating domains typically exhibits a power law dependence on time, characterized by a growth exponent.
\begin{equation}
    l(t)\sim t^{1/z}
    \label{eq:expo}
\end{equation}
where $1/z$ is the growth exponent. FIG.\ref{fig:10}(c) shows log-log plot of characteristic length \(l(t)\) versus time \(t\) for two different quench temperatures \(T_h^*=25\) and \(40\) in MD simulation. The characteristic length \(l(t)\) grows algebraically, with the growth exponent  \(1/z\) approximately equal to \(1/3\) at late times. The values of  \(1/z=0.32\) and \(0.31\) give best fitting for quench temperatures \(T_h^*=25\) and \(40\) respectively. The growth exponent of \(1/3\) is widely reported in passive systems with conserved order parameter referred as Lifshitz-Slyozov\cite{LIFSHITZ196135} exponent.  \\
 Additionally, the plot of correlation length, $l(t)$, vs. time, $t$, obtained from CG model for different activity is shown in FIG.\ref{fig:3}(a). The late time behavior of correlation length, $l(t)$ aligns well with growth exponent $1/z = 1/3 \approx 0.33$. The behavior of $l(t)$ is consistent over the entire range of activity considered in this study. In the context of coarsening systems, previous studies \cite{shinozaki1992spinodal,koga1993late} have established that, when the domain sizes are  $\gtrsim 25 \%$ of the linear dimension of the system, $L$, the domain growth is strongly influenced by finite size effects. In our simulations, we observe that these finite size effects lead to a pronounced slowing down of growth at very late times. To ensure that the observed growth law is not an artifact of the finite size effect we perform the finite size scaling analysis of the data set. Here we present the finite size scaling analysis for  activity $\chi = 2.50$. This analysis yields consistent results for the other values of $\chi$.\\
 
\underline{{\em Finite size analysis for growth exponent:}} The time, $t_{eq}$, after which $l(t)$ saturates varies with the system size, $L$, as: $t_{eq} \sim L^{\beta}$. 
We propose the scaling form of $l(t)$ as 
\begin{equation}
    l(t,L) = t^{1/z} f(t/t_{eq})
    \label{eq:8}
\end{equation}
The properties of the scaling function, $f(x)$, are as follows: (i) {for $x \rightarrow 0$, $f(x) = $ constant i.e. $l(t) \sim t^{1/z}$} and (ii) for $x \rightarrow \infty$, $f(x) \sim x^{-1/z} $ i.e. $l(t) = $ independent of $t$. Hence we can write,
\begin{align}
        &l(t,L) = t^{1/z} f(L^{-\beta} t) \notag \\
        &\Rightarrow l^{z}(t,L) = t f^{z}(L^{-\beta} t) \notag \\
        &\Rightarrow t^{-1}l^{z}(t,L) = f^{z}(L^{-\beta} t) \notag \\
        \label{eq:9}
\end{align}

As shown in FIG.\ref{fig:3}(b), the plot of $t_{eq}$ $vs.$ system size, $L$ gives $\beta = 3.3092$. Next we plot $t^{-1}l^{z}(t,L)$ $vs.$ $L^{-\beta} t$ for different system sizes, $L$. Only appropriate value of $z$ will give good collapse of data for various system sizes. In FIG.\ref{fig:3}(c) $\&$ (d), data collapse is shown for $1/z = 0.33$, and $1/z = 0.25$. For $1/z = 0.33$ we observe good data collapse, whereas for $1/z = 0.25$ there is no data collapse. This establishes the value of the growth exponent $1/z = 1/3$ in the CG model which matches with microscopic results.


\begin{figure*}[hbtp!]
\centering
\subfigure[]{
  \includegraphics[width=0.23\textwidth]{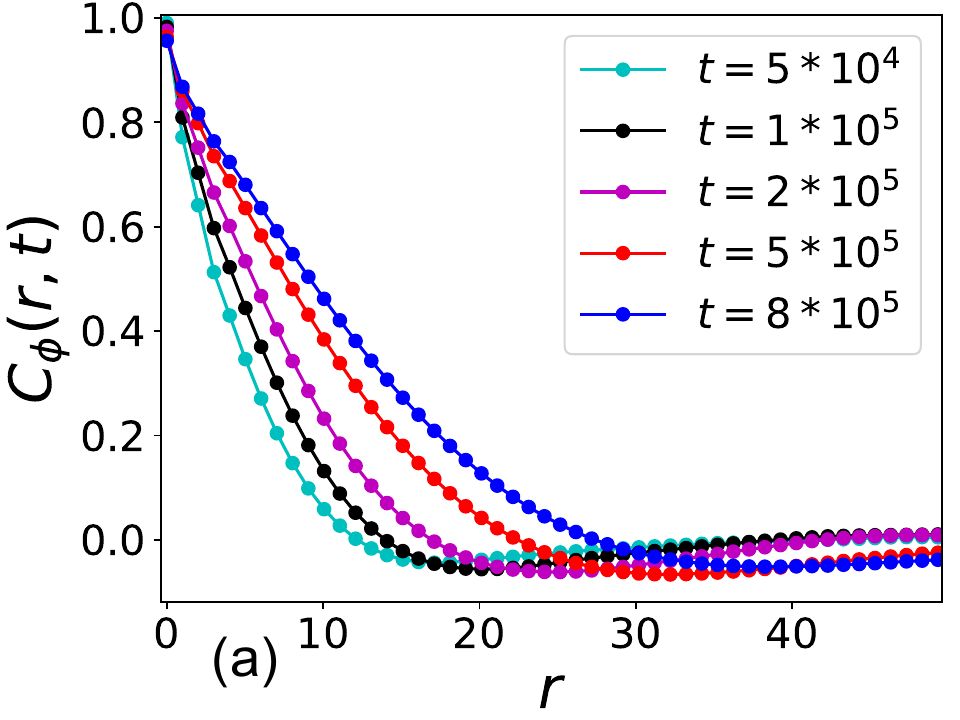}
}
\subfigure[]{
  \includegraphics[width=0.23\textwidth]{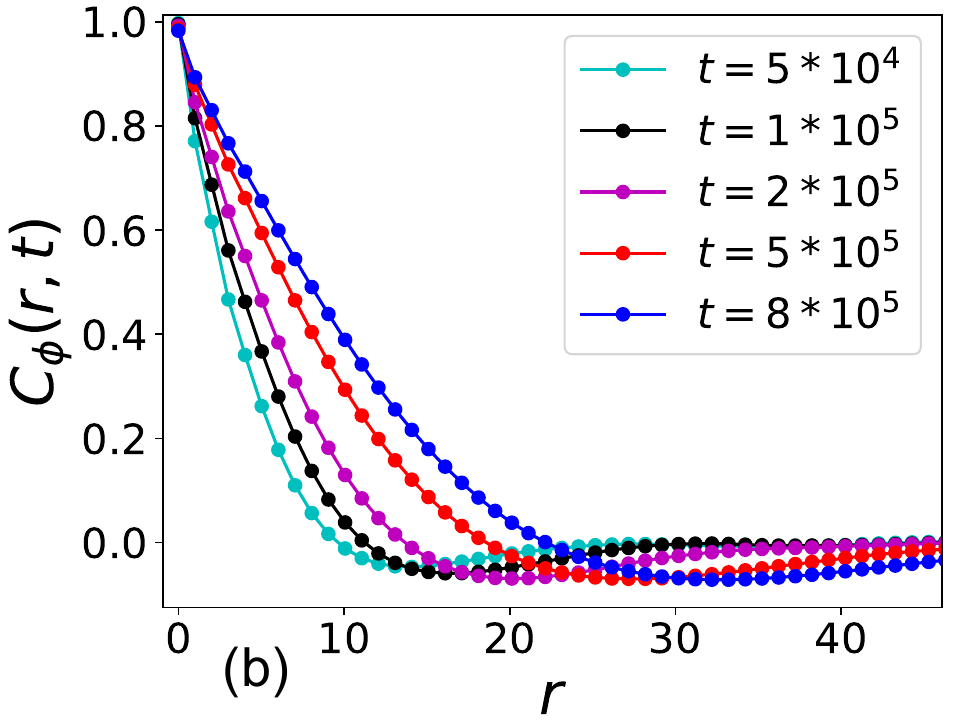}
}
\subfigure[]{
\includegraphics[width=0.23\textwidth]{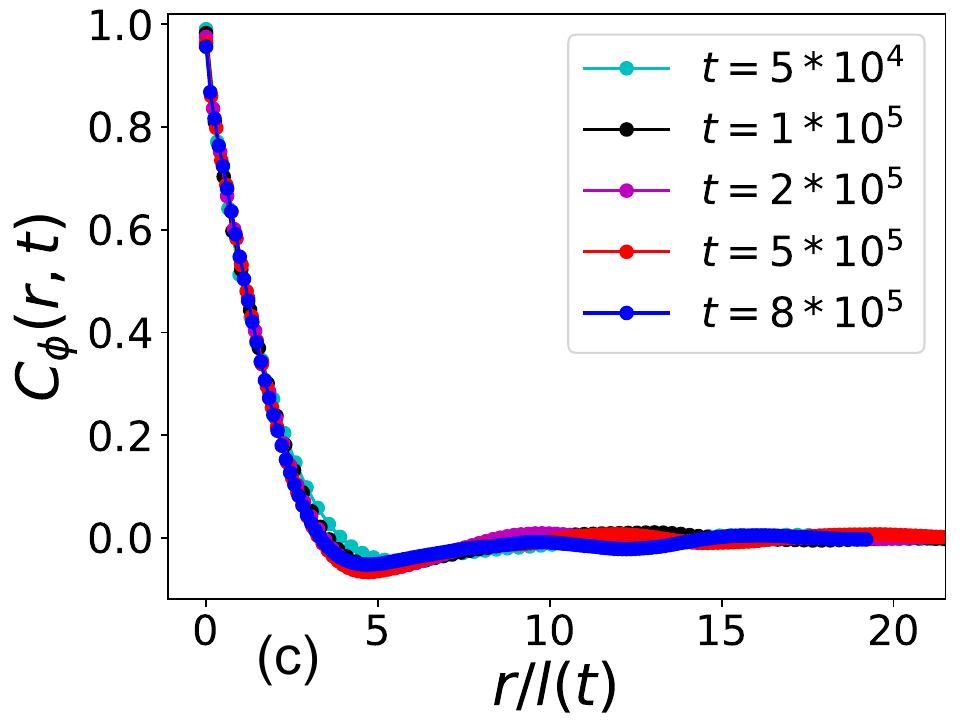}
}
\subfigure[]{
  \includegraphics[width=0.23\textwidth]{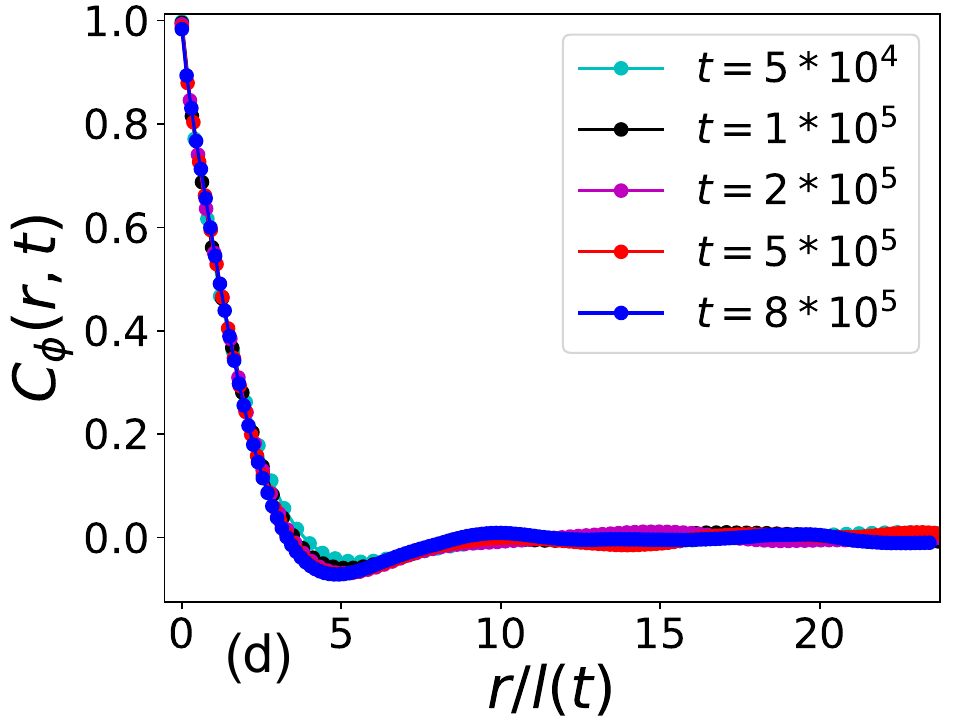}
}%

\caption{(a) and (b) Plots of two-point spatial correlation function of \(\phi\), \(C_{\phi}(r,t)\) as function of distance between the points \(r\) at density \(\rho^*=0.8\) for quench temperature \(T_h^*=25\) and  \(T_h^*=40\) respectively, at different instants of time in 2D. The correlation function \(C_{\phi}(r,t)\) decays slowly as time progresses. (c) and (d) Plots of two-point spatial correlation function \(C_{\phi}(r,t)\) as function of distance \(r\) rescaled by the characteristic length \(l(t)\) at density \(\rho^*=0.8\) for quench temperature \(T_h^*=25\) and  \(T_h^*=40\) respectively at different instants of time. The correlation functions at different times overlap onto a master curve implying the self-similar nature of domain growth.  }
\label{fig:cor_2d}
\end{figure*}

\begin{figure*}[hbtp!]
\centering

\subfigure[]{
  \includegraphics[width=0.31\textwidth]{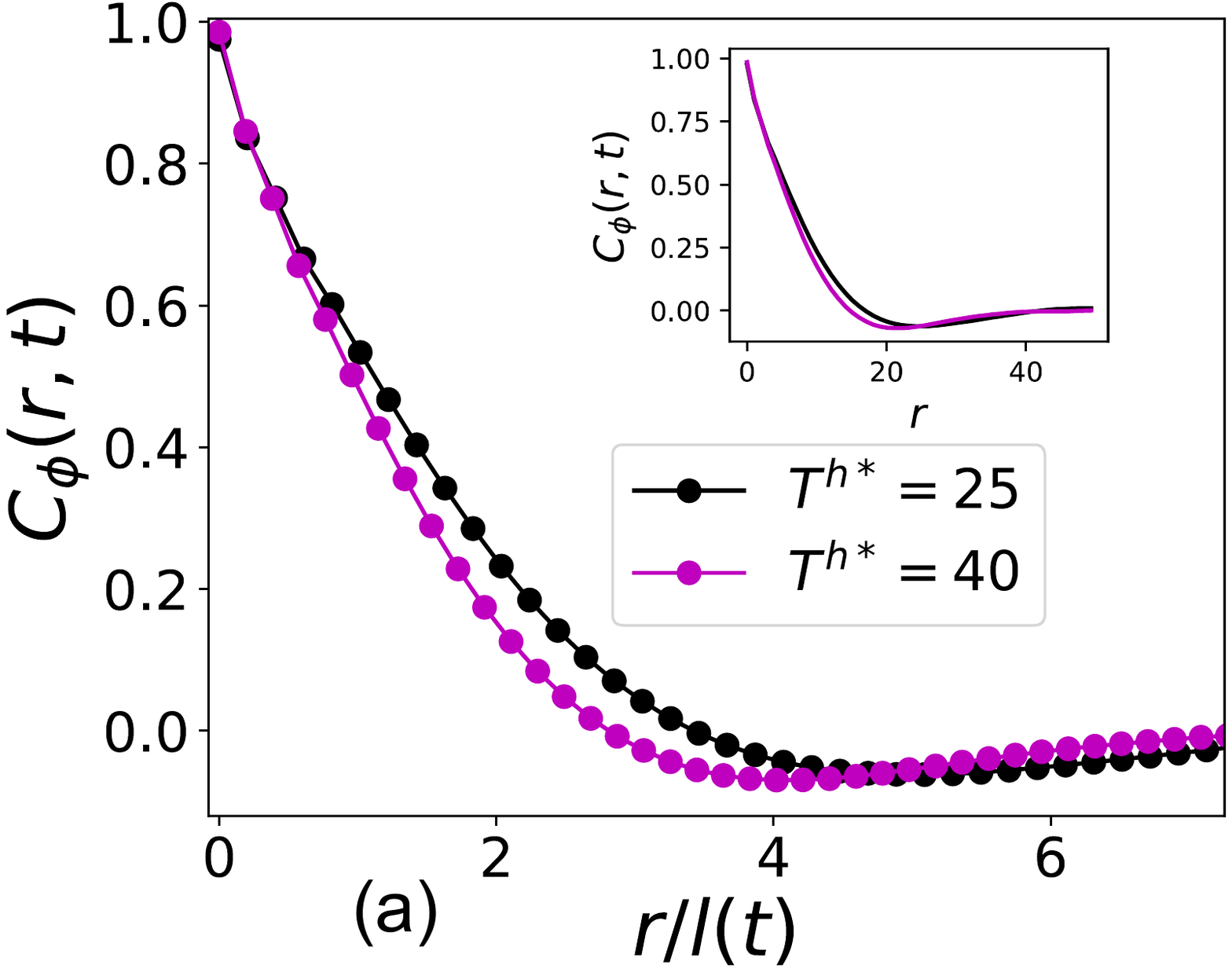}
}
\subfigure[]{
  \includegraphics[width=0.31\textwidth]{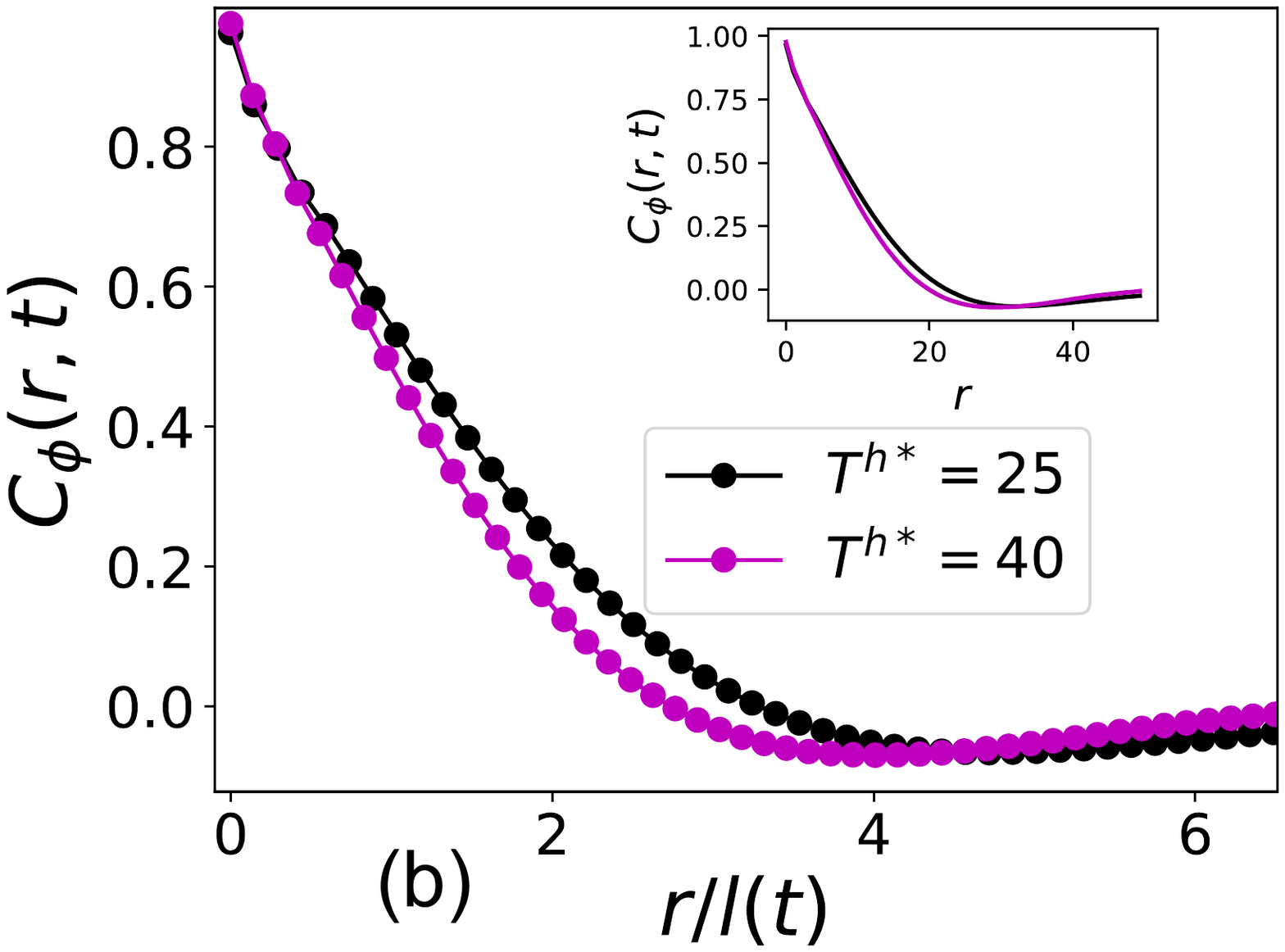}
}
\subfigure[]{
  \includegraphics[width=0.31\textwidth]{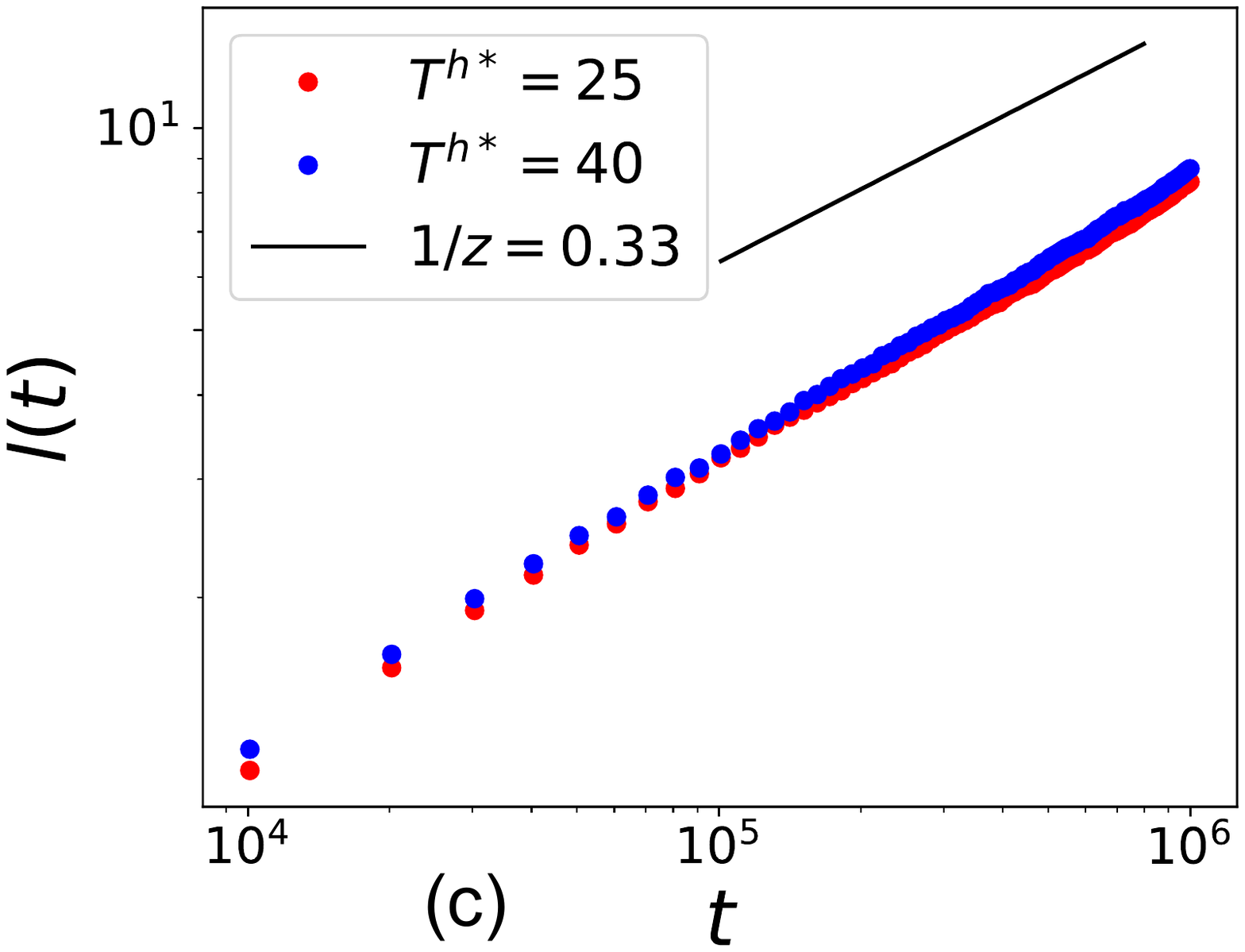}

}

\caption{(a) and (b) Plots illustrate static characteristic of \(C_{\phi}(r,t)\) as a function of rescaled distance \(r/l(t)\)  for different quench temperature at time \(t=2*10^5\) and \(t=5*10^5\), respectively, in 2D. The insets show \(C_{\phi}(r,t)\) as function of distance \(r\) only. The plots reveal that static scaling in 2D is not as robust as in 3D.  (c) Log-Log plot of characteristic length \(l(t)\) versus time \(t\) shows algebraic growth of \(l(t)\). We see that the growth exponent \(1/z\) has a value close to 1/3 at late times.  }
\label{fig:static_2d}
\end{figure*}

\begin{figure*}[hbtp!]
\centering
 \includegraphics[width=0.6\textwidth]{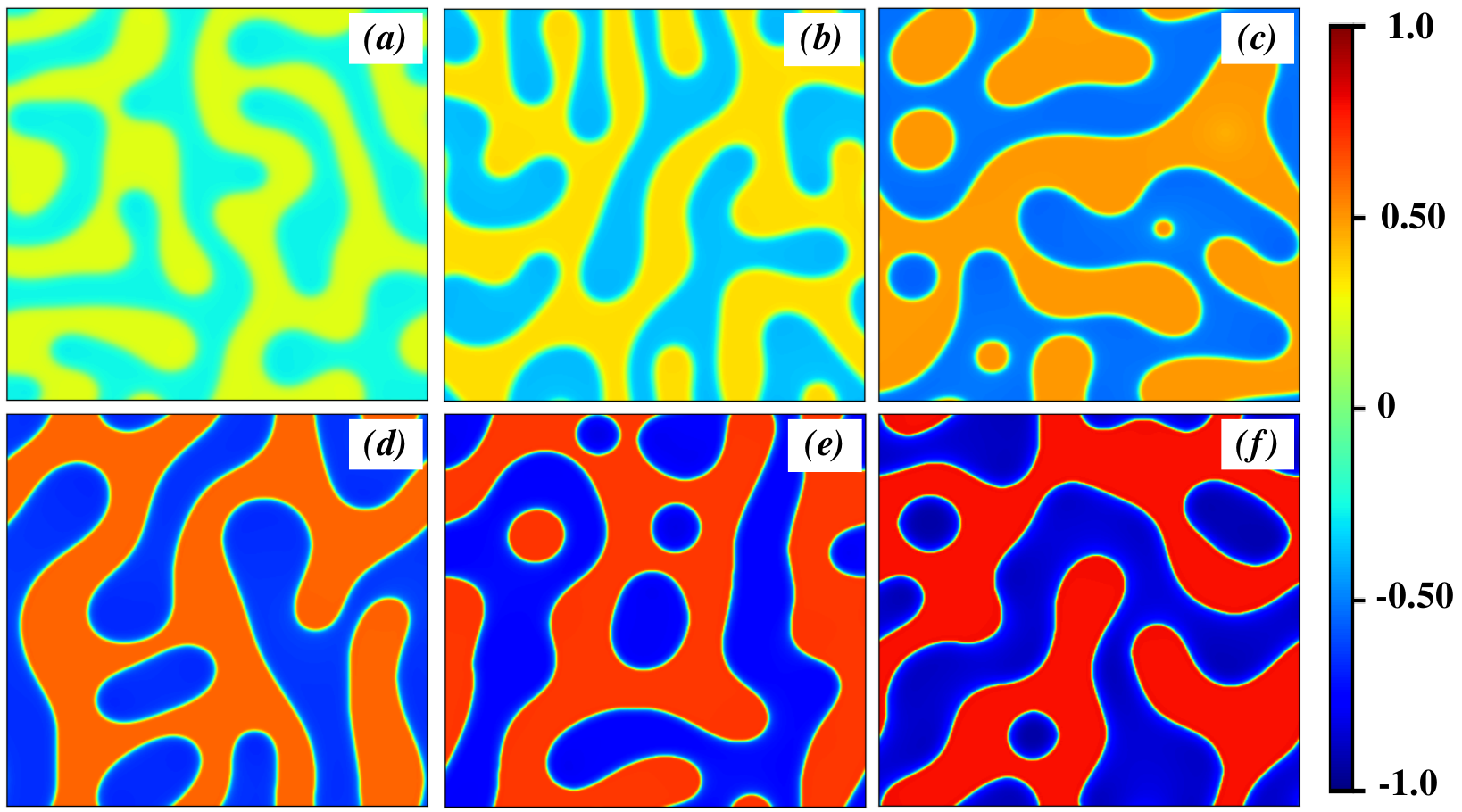}
 \includegraphics[width=0.3\textwidth]{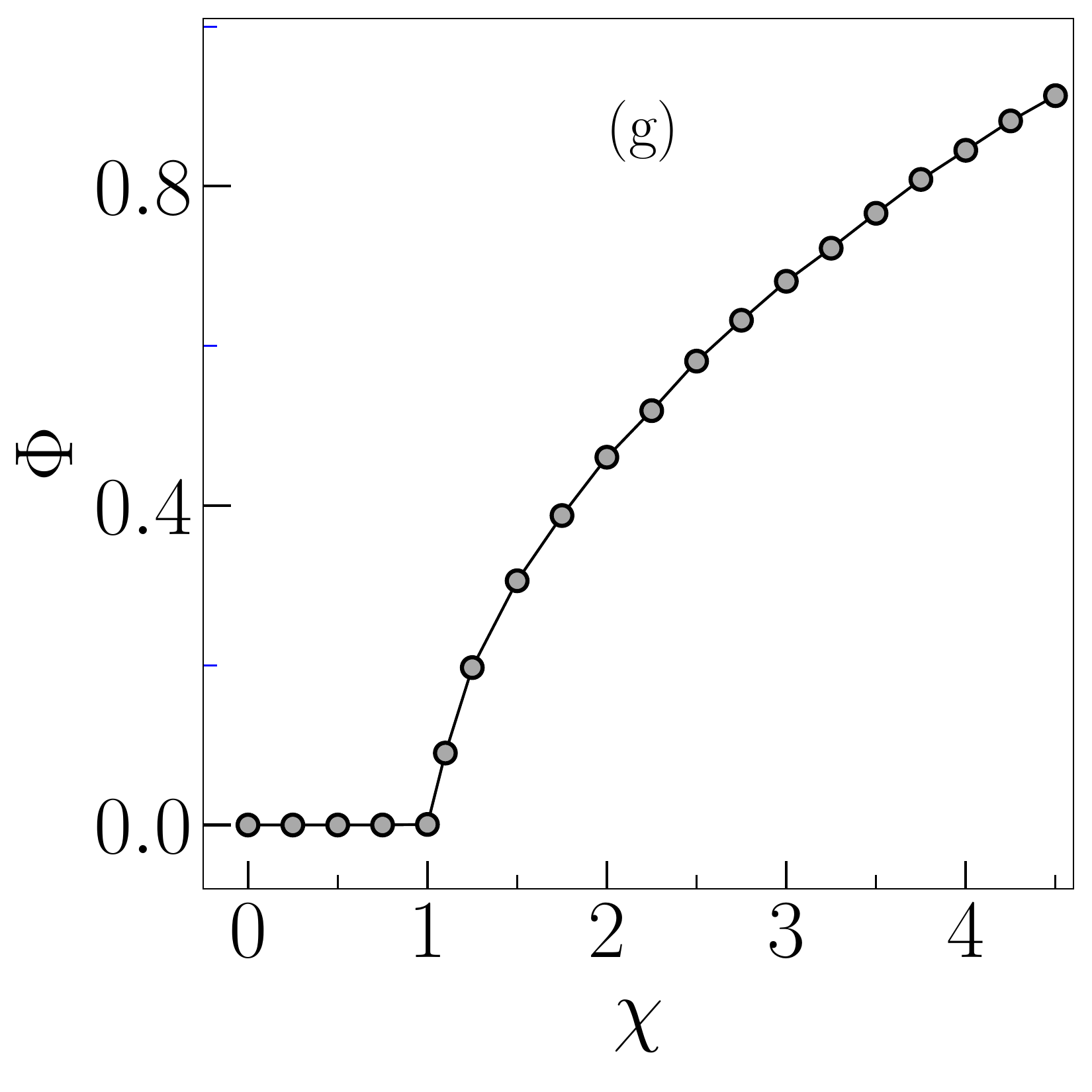}

\caption{The figure shows the phase separation in the critical mixture in 2D for different activity. The series of snapshots (a)-(f) show the plot of $\phi$ according to the colorbar at late time ($t = 10^4$) for different values of $\chi$: (a) $\chi = 1.25$, (b) $\chi = 1.50$, (c) $\chi = 2.00$, (d) $\chi = 2.50$, (e) $\chi = 3.00$, and (f) $\chi = 3.50$. The subplot (g) shows the variation of PSOP, $\Phi$, with activity, $\chi$. System size, $L = 256$  }
\label{fig:2d_crit_psop}
\end{figure*}


\subsection{2-TIPS in 2-dimension}\label{sec:md_2d}

\underline{\textit{\text{Kinetics of phase separation}}}: Instantaneous snapshots of the 2D binary mixture in MD simulations comprising of 80,000 particles undergoing 2-TIPS shown in FIG.\ref{fig:11}(a), capture the temporal evolution of hot and cold domains, following the quenching of the system to \(T_h^*=25\). Akin to phase separation kinetics in 3D, the hot and cold particles phase separate by the formation of dense (cold particle rich) and dilute (hot particle rich) bi-continuous domains, which grow in size over time. However, we see increased trapping of hot particles in the cold domains compared to the 3D system for reasons discussed in appendix \ref{app:premd}. The formation of micro-domains of hot particles in the cold regions reduces its effective domain size. By utilizing the local density field \(\rho(r)\) to distinguish the hot and cold domains, as detailed in Sec.\ref{sec:md_3d}, we overcome the issue of underestimating the domain size of cold regions. From the local density field \(\rho(r)\), we obtain the order parameter field \(\phi(r,t)\) by using the same methodology followed in the 3D system. FIG.\ref{fig:11}(b) illustrates the order parameter field \(\phi(r,t)\) at different instances of time for quench temperature \(T_h^*=25\). The dark blue regions indicate domains with the majority of cold particles (\(\phi(r)=1\)), and dark red regions correspond to domains rich in hot particles (\(\phi(r)=-1\)).\\

\begin{figure*}[hbtp!]
\centering

 \includegraphics[width=0.8\textwidth]{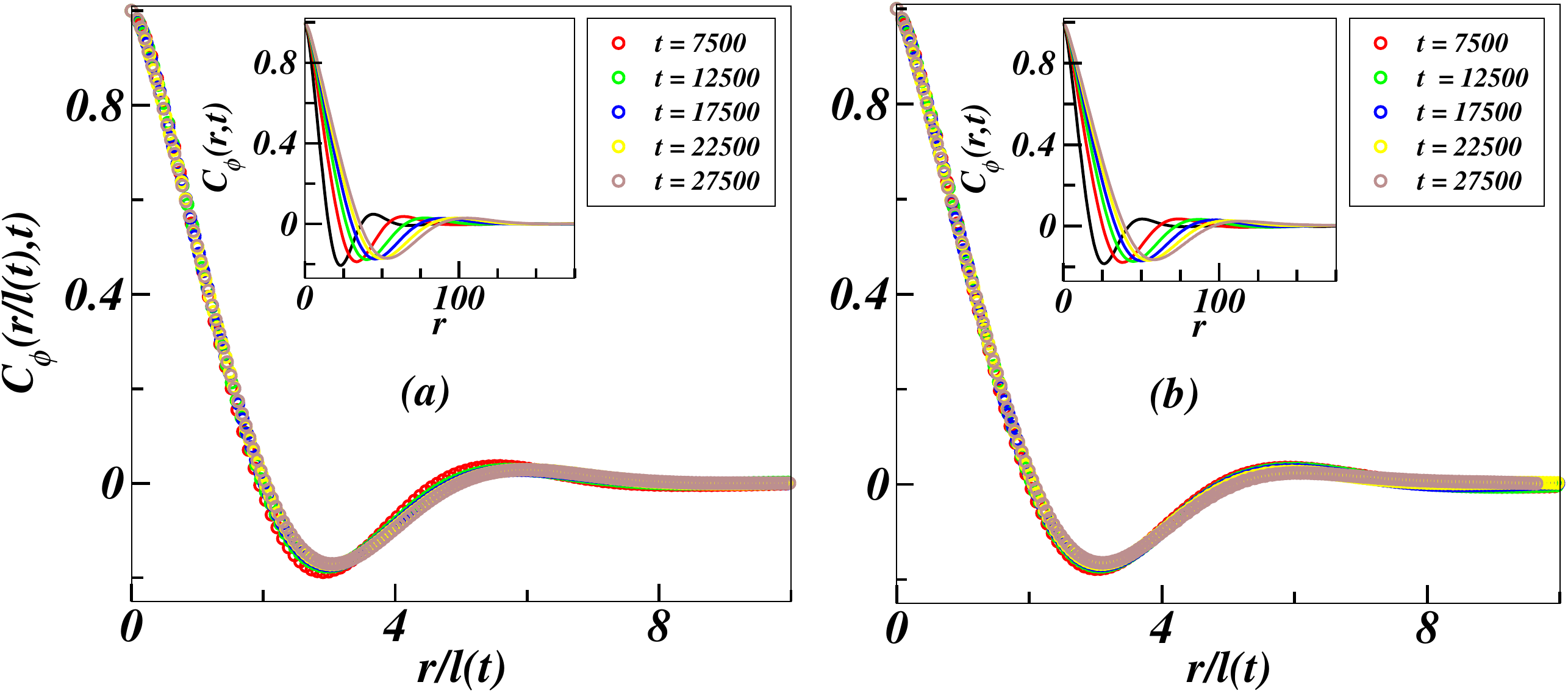}
 \includegraphics[width=0.8\textwidth]{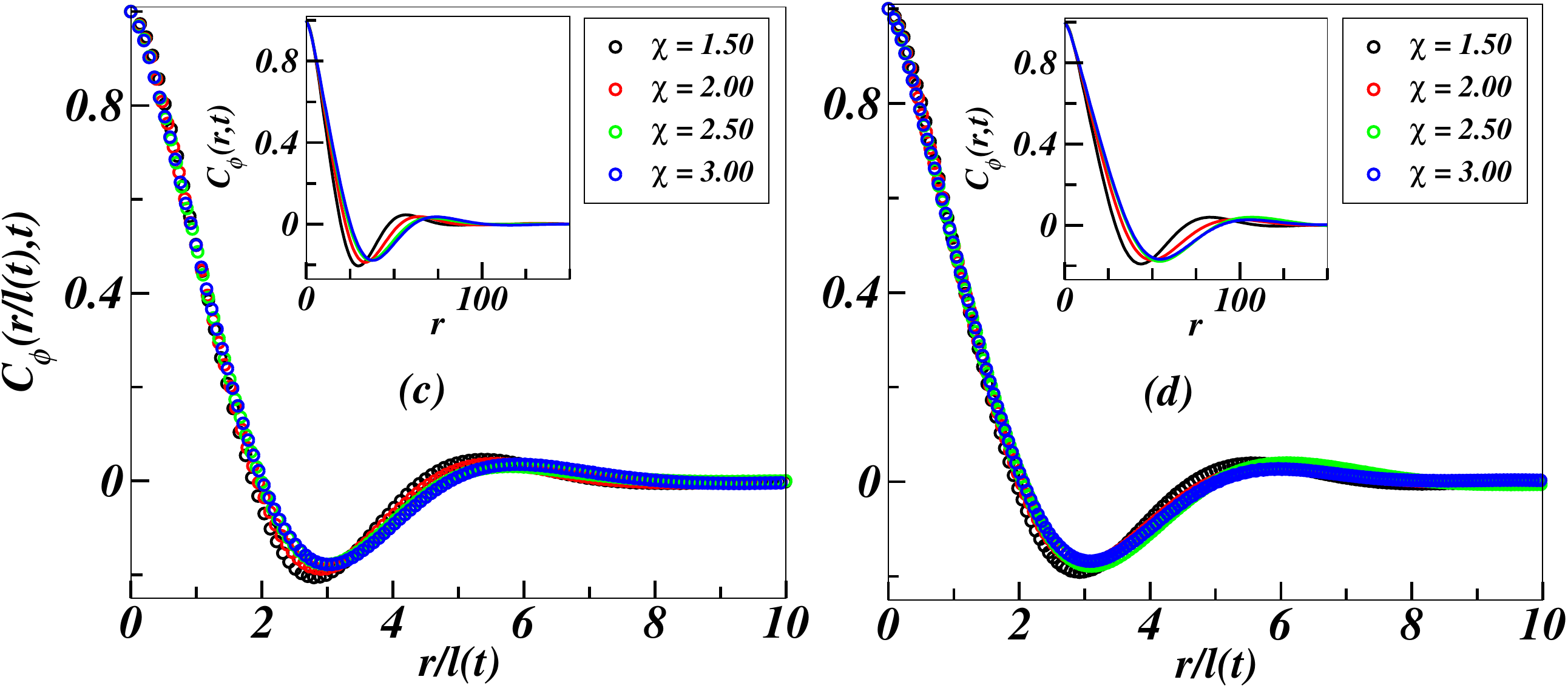}

\caption{The figure depicts the behavior of the two point correlation function, $C_{\phi}(r,t)$ for the critical mixture in 2D. The color code shown in legends applies for the main plot as well as in all the subplots (a)-(d). The subplots (a)-(b) show the dynamical scaling behavior of $C_{\phi}(r,t)$ for two different activity: (a) $\chi = 2.00$, and (b) $\chi = 3.00$. The insets show the plot of the correlation function at different times, and the main plots show the scaling collapse of the correlation function when the distance, $r$, is scaled by the correlation length, $l(t)$. The subplots (c)-(d) show the static scaling behavior of $C_{\phi}(r,t)$ with respect to activity at two different times: (c) $t = 7.5 \times 10^3$, and (d) $t = 2.25 \times 10^4$. The insets show the plot of $C_{\phi}(r,t)$, and the main plots show the scaling collapse of $C_{\phi}(r,t)$ for different $\chi$ when distance, $r$, is scaled by correlation length, $l(t)$. System size, $L = 512$}
\label{fig:2D_crit_corr}
\end{figure*}

Again, to quantify the domain structures in 2D, we calculate the two-point spatial correlation function of  \(\phi\), $C_{\phi}(r,t)$ defined in Eq.\ref{eq:cor}. FIG.\ref{fig:cor_2d}(a-b) depict the plot of the correlation function $C_{\phi}(r,t)$  as a function of distance \(r\) between two points for 2D simulation for quench temperatures \(T_h^*=25\) and  \(T_h^*=40\). The decay in correlation function $C_{\phi}(r,t)$ becomes progressively slower with time, indicating growth in the domain size. When the distance between the two points \(r\) is rescaled by the characteristic length \(l(t)\), the correlation functions $C_{\phi}(r,t)$ at different times collapse onto a master curve as shown in figures \ref{fig:cor_2d}(c-d). This shows that even in 2D systems, the domain morphologies of the phase-separating binary mixture at different instances of time are statistically similar. FIG.\ref{fig:static_2d}(a-b) shows the plot of \(C_{\phi}(r,t)\) for two different quench temperatures at times \(t=2 \times 10^5\) and \(t=5 \times 10^5\), respectively.  However, as demonstrated in FIG.\ref{fig:static_2d}(a-b), unlike 2-TIPS in 3D, the static scaling of $C_{\phi}(r,t)$ is not very robust. \\
To further clarify the kinetics of phase separation and growth exponent we turn to the CG model of 2-TIPS. In FIG.\ref{fig:2Dcrit_timeseries}, we show the time evolution of the phase separation variable $\phi$ in a critical mixture in 2D starting from an initial homogeneous configuration. At very early time, the hot and cold particles in the system are mixed leading to homogeneous distribution of $\phi$ in the entire system. However, as time elapses, the cold particles start to form bicontinuous clusters in the sea of disordered hot particles leading to large positive and negative values of $\phi$. In the snapshots, $\phi > 0$ and $\phi < 0$ represents the regions occupied predominantly by hot and cold particles, respectively. The size of the structures increases with time. However, the time required to reach a complete phase separated state scales with the system size.\\
The FIG.\ref{fig:2d_crit_psop} depicts the dependence of phase separation in the system on the activity, $\chi$. The snapshots (a)-(f) show the enhancement of phase separation in the system with increase of activity. A quantitative measure of local phase separation between hot and cold particles can be obtained by the color contrast between the $\phi ({\bf r})>0$ and $\phi({\bf r})<0$ regions. The snapshots clearly show that with an increase in activity, the system exhibits enhanced phase separation between the hot and cold particles with well-defined interfaces. The subplot (g) showcases the variation of PSOP, $\Phi$, with activity, $\chi$, which provides a measure of phase separation in the system on a global scale.\\ 
To study the kinetics of domain growth in the mixture, we calculate the spatial two point correlation function of $\phi$, $C_{\phi}(r,t)$, given by Eq.\ref{eq:cor}. The growth of domains over time is reflected in the slower decay of $C_{\phi}(r,t)$ as time progresses. FIG.\ref{fig:2D_crit_corr} showcases the scaling properties of the two point correlation function. In FIG.\ref{fig:2D_crit_corr}(a)-(b) we show the dynamic scaling of $\phi$, $C_{\phi}(r,t)$ at two different activity. The correlation function at different times collapses onto a single master curve when the distance, $r$, is scaled by the characteristic length scale, $l(t)$. This shows the statistically similar morphology of domains at different stages during the domain growth. However, unlike the 3D critical mixture, in the 2D critical mixture we did not observe a good static scaling of the correlation function with respect to activity as shown by FIG.\ref{fig:2D_crit_corr}(c)-(d). \\

\begin{figure*}[hbtp!]
\centering
  \includegraphics[width=0.35\textwidth]{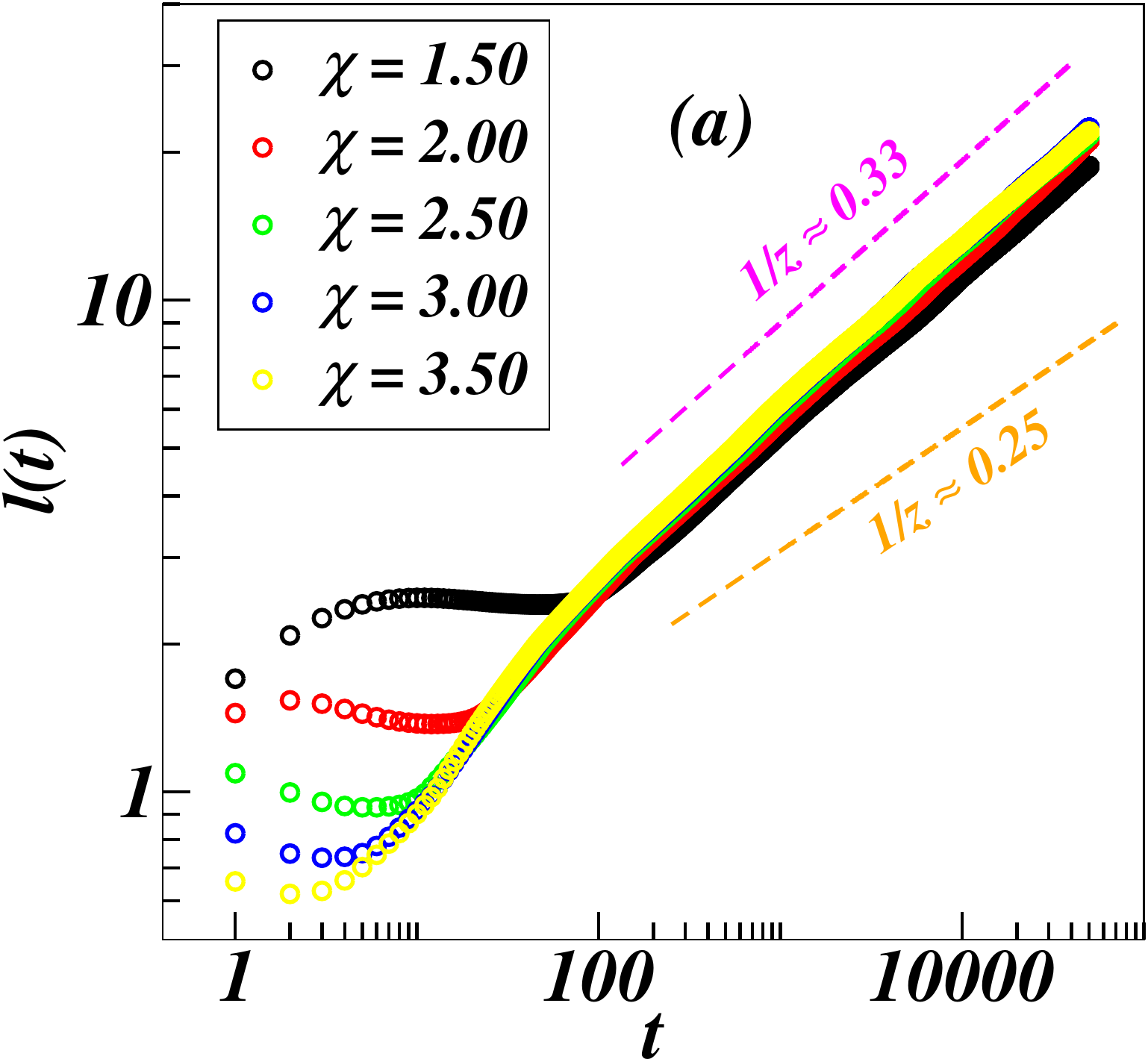}
  \hspace{1 cm}
  \includegraphics[width=0.51\textwidth]{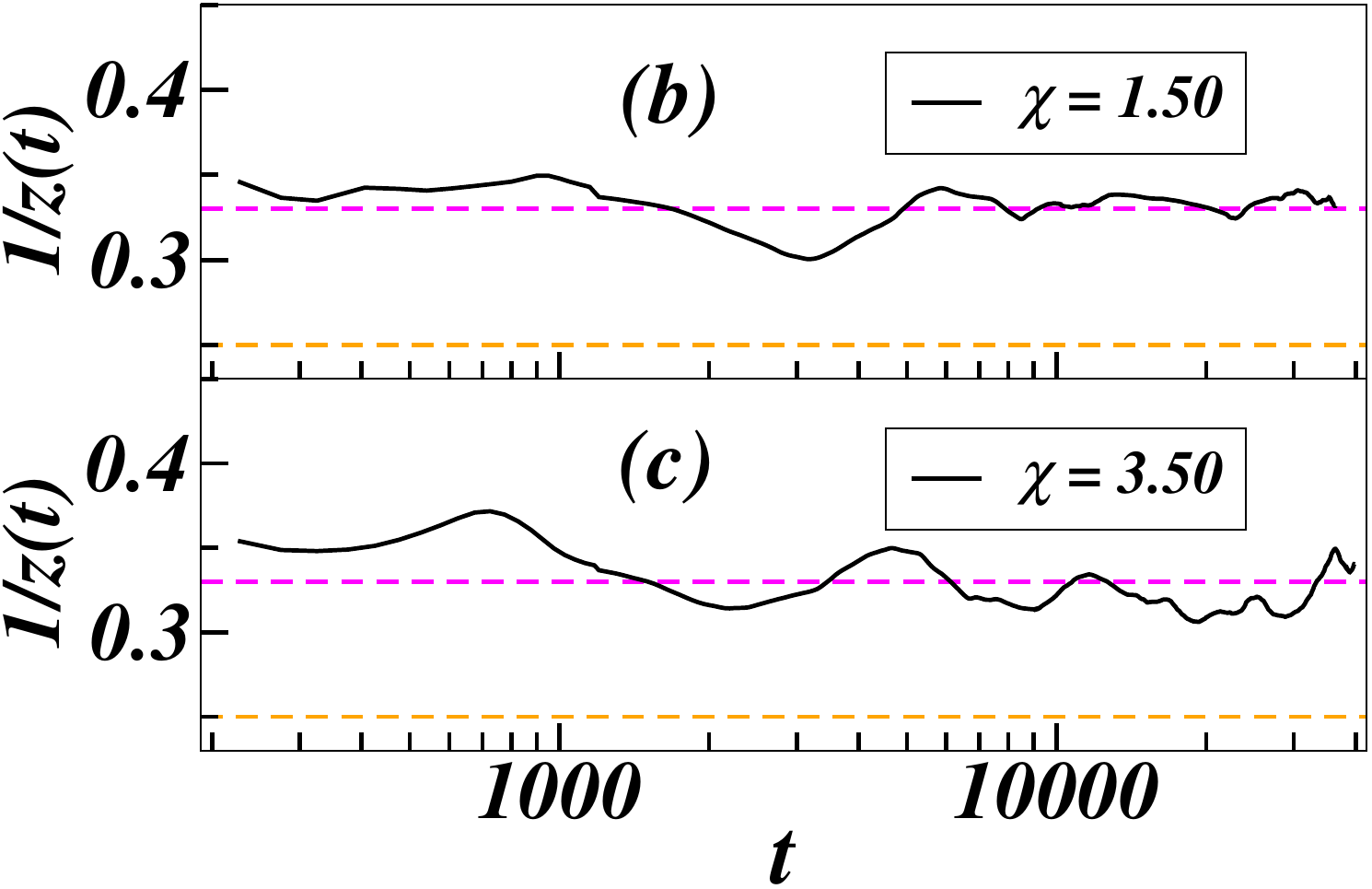}

\caption{The figure depicts the characteristics of the domain growth in 2D. The subplot(a) shows the variation of the characteristics length scale, $l(t)$, with time, (t), for different activity on a log-log scale. The dashed line labelled $1/z = 0.33$ denotes growth law $l(t) \sim t^{1/3}$. The subplots (b) and (c) shows the variation of the effective exponent $1/z(t)$ with time for two different values of activity: (b) $\chi = 1.50$, and (c) $\chi = 3.50$. The dashed lines in magenta and cyan denote the effective exponent $1/z(t) = 0.33$ and $1/z(t) = 0.25$. System size, $L = 512$ }
\label{fig:2D_len_scale}
\end{figure*}

\underline{{\em Growth exponent}}: FIG.\ref{fig:static_2d}(c) reveals log-log plot of characteristic length \(l(t)\) versus time \(t\) for quench temperatures \(T_h^*=25\) and \(40\) in MD simulations. Again, we see algebraic growth of characteristic length \(l(t)\) with time. We find that irrespective of the dimensionality of the system, the growth exponent \(1/z\) associated with 2-TIPS at high density of \(\rho^*=0.8\) is approximately equal to \(1/3\) at late times. The values of \(1/z=0.35\) and \(0.34\) give best fitting for quench temperatures \(T_h^*=25\) and \(40\) respectively. Further the plot of $l(t)$ vs. $t$ obtained from CG simulation is shown in FIG.\ref{fig:2D_len_scale}(a). $l(t)$ varies algebraically with $t$, expressed by Eq.\ref{eq:expo}. As shown in FIG.\ref{fig:2D_len_scale}(a), the growth exponent is consistent with $1/3$ across all the values of $\chi$.  To further validate this, we calculate the effective growth exponent, $1/z(t)$, as a function of $t$ as follows
\begin{equation}
    \frac{1}{z(t)} = \frac{d \{ln(l(t))\}}{d \{ln(t)\}}
    \label{eq:10}
\end{equation}
FIG.\ref{fig:2D_len_scale}(b)-(c) showcase the variation of $1/z(t)$ as a function of time, $t$, for two different values of $\chi$. As is typical with effective exponents, these plots exhibit significant fluctuations. However, the overall trend is clear, showing that in both cases, the value of $1/z(t)$ fluctuates around $1/3$ at late times.

{\section{Discussion}\label{secIV}}
In this study, we probe the phase separation kinetics of the binary mixture of hot and cold particles undergoing 2-TIPS by MD simulation and a CG model in 3D and 2D.  Several features distinguish 2-TIPS from equilibrium phase separation, the most prominent being the difference in the interactions between the cold and hot species. Specifically, interactions between a cold and a hot particle cause the cold particle to accelerates while the hot particle slows down. This results in distinct trends in the effective temperatures of the cold and hot species, deviating from their respective original temperatures. Our coarse-grained model incorporates this inherent feature. \\
In the CG model, the hot and the cold particles are represented by two coupled free energies, with hot particles maintained above the critical temperature and cold particles below it. The coupling between the two species is introduced in such a way that it modifies their effective temperatures according to $T_h^* > T_h^{eff*}> T_c^{eff*} > T_c^*$. The introduction of such a coupled free energy is one of the main idea of this work.\\
Both the MD simulations as well as the CG model reveal that phase separation kinetics in 2-TIPS at high density is similar to spinodal decomposition in passive/equilibrium systems, where the hot and cold particles phase separate by the formation and growth of bi-continuous domains.
The phase separation increases with increase of activity. In the kinetic regime, the system exhibits both static and dynamic scaling, indicating a statistically self-similar domain morphology over time and across varying activity levels. Hence, the non-equilibrium kinetics of 2-TIPS share many of the universal features exhibited by passive systems like dynamic scaling and algebraic growth in domain size.\\
The growth exponent for 2-TIPS from both 3D and 2D simulations is close to 1/3, commonly referred to as the Lifshitz-Slyozov exponent. Further, the phase separation in 2-TIPS has facile analogies with liquid vapor condensation \cite{10.1063/1.472839} in equilibrium and MIPS in non-equilibrium, with the coexistence of dense and dilute phases being the common feature among them. The underlying dynamics driving phase separation in the MIPS and 2-TIPS systems are fundamentally different.
In MIPS, the effective local density of particles, results in the reduced motility of particles and hence the trapping, which is responsible for the phase separation.
In 2-TIPS,  the activity originates from inter-species interactions between the cold and hot species. During these interactions, energy transfer from hot particles to cold particles increases the effective temperature of the cold particles while decreasing that of the hot particles. The temperature difference between the particles leads to density difference, causing the cold particles to be caged by the hot particles.  Both systems are inherently non-equilibrium and possess non-zero interface current \cite{PhysRevE.107.024701}, which increases with increasing activity parameter as shown in FIG.\ref{fig:current_snap} and FIG.\ref{fig:current_mag}.
Even though the underlying mechanism and interactions vary greatly, Lifshitz Slyozov domain growth law is also observed in passive systems undergoing condensation \cite{10.1063/1.472839}, ABP's undergoing MIPS \cite{D0SM01762K,PhysRevE.108.024609} and in 2-TIPS as demonstrated by our results. \\
The findings from MD and CG modeling show good agreement and indicate that the universality in phase separation kinetics also extends to 2-TIPS, a non-equilibrium phase separation phenomenon, beyond the scope of equilibrium systems. In the future, we would like to continue our investigations in low-density limit where hydrodynamic effects could play a significant role in the kinetics of 2-TIPS. We would also like to study kinetics of 2-TIPS in liquid crystals to see the effect of anisotropy of the particles on the kinetics.\\

\section{Acknowledgement}
N.V. and J.M. thank MoE, India, and CSIR, India, respectively, for the fellowship. P.S.M. and S.M. thank PARAM Shivay for computational facility under the National Supercomputing Mission, Government of India at the Indian Institute of Technology, Varanasi and the computational facility at I.I.T. (BHU) Varanasi. P.S.M.  thanks UGC for research fellowship. S.M. thanks DST, SERB (INDIA), Project No.: CRG/2021/006945, MTR/2021/000438  for financial support. We also thank Profs. Sumantra Sarkar and Subir K Das for insightful discussions.\\

\newpage
\onecolumngrid
\appendix

\section{Previous Results from MD study}\label{app:premd}

In our previous studies \cite{chari2019scalar,D3SM00796K}, we have thoroughly examined the non-equilibrium steady states of the binary mixture hot and cold LJ particles undergoing 2-TIPS, using MD simulations in 3D as well as 2D. Starting from a homogenous configuration where both hot and cold particles were assigned the same temperature \(T_c^*=T_h^*=2\), the temperature of hot particles was raised to \(T_h^*= 80\) in increments of 5.\\
The simulations revealed that the hot and cold LJ particles undergo phase separation when the activity \(\chi\) exceeds a density-dependent critical value \(\chi_{crit}\). The FIG.\ref{fig:snap_ss}  shows the snapshots of the instantaneous configurations of the binary mixture in its initial mixed state (\(T_h^*=2\)) and the final phase separated state (\(T_h^*=80\)) at density \(\rho^*=0.8\) in 3D and 2D. The high kinetic pressure of the hot particles forces the cold particles to phase separate by the formation of dense clusters with crystalline order. Once the binary mixture of hot and cold particles reached the steady state, the extent of phase separation was quantified by an order parameter \(\phi\).  The simulation volume/area was divided into sub-cells, and the local order parameter \(\phi\) was obtained by taking the average of the normalized number difference between hot and cold particles \(\frac{|n_h-n_c|}{(n_h+n_c)}\) over all the sub-cells and steady state configurations. FIG.\ref{fig:op_ss}(a) and (b) illustrate the plots of order parameter \(\phi\) versus activity \(\chi\) at different densities for 3D and 2D systems, respectively. We see that the order parameter \(\phi\) increases with activity \(\chi\) at all densities, implying enhanced phase separation at higher activity. One notable difference between 3D and 2D systems is that the order parameter \(\phi\) increases with density in 3D, whereas an opposite trend is observed in 2D. This reduction in order parameter \(\phi\) with density is attributed to the enhanced trapping of hot particles in the dense cold domains in 2D \cite{D3SM00796K}. The reduction in dimension diminishes the domain interface from a surface to a curve leading to decreased pathways for the escape of hot particles entrapped in the cold domains. However, the size of the largest cluster in the phase separated 2D binary mixture (see ref \cite{D3SM00796K}) increases with density, implying enhancement in phase separation with density similar to the 3D case. The phase diagram for 2-TIPS of LJ particles in 3D in the activity \(\chi\) - density \(\rho\) phase space is depicted in FIG.\ref{fig:op_ss}(c). In the phase diagram, the purple color denotes the phase-separated region, whereas the white color indicates the mixed-state regime. The grey color is the critical region encompassing the critical line. The phase diagram of 2-TIPS in 2D is qualitatively similar to 3D with critical activity \(\chi=3.59, 6.08, 8.68\) and \(11.18\) corresponding to densities \(\rho^*=0.8, 0.5, 0.2\) and \(0.1\) respectively. Readers are referred to our previous articles \cite{chari2019scalar,D3SM00796K} for further details.
\setcounter{figure}{0}
\renewcommand{\thefigure}{A\arabic{figure}}

\FloatBarrier
 \begin{figure*}[htb]
 \centering 
 \includegraphics[width=0.9\textwidth]{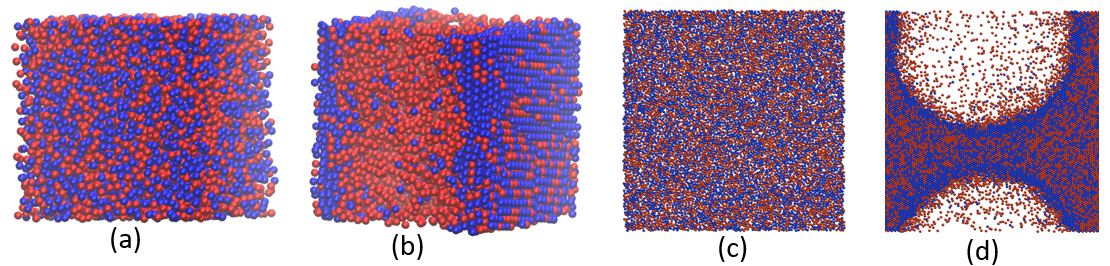}
 \caption{ (a) and (b) show instantaneous snapshots of the 3D system containing 8000 hot and cold particles in the initial state with \(T_h^*=2\) and the final phase separated state with \(T_h^*=80\) respectively at density \(\rho^*=0.8\).  (c) and (d) show instantaneous snapshots of the 2D system containing 8000 hot and cold particles in the initial state with \(T_h^*=2\) and the final phase separated state with \(T_h^*=80\) respectively at density \(\rho^*=0.8\). Starting from a well-mixed state, the binary mixture undergoes phase separation when the temperature of the hot particles is increased to \(T_h^*=80\) in 3D as well as 2D. }
 \label{fig:snap_ss}
 \end{figure*}
 \FloatBarrier

\FloatBarrier
\begin{figure*}[htb]
\centering
\subfigure[]{
  \includegraphics[width=0.31\textwidth]{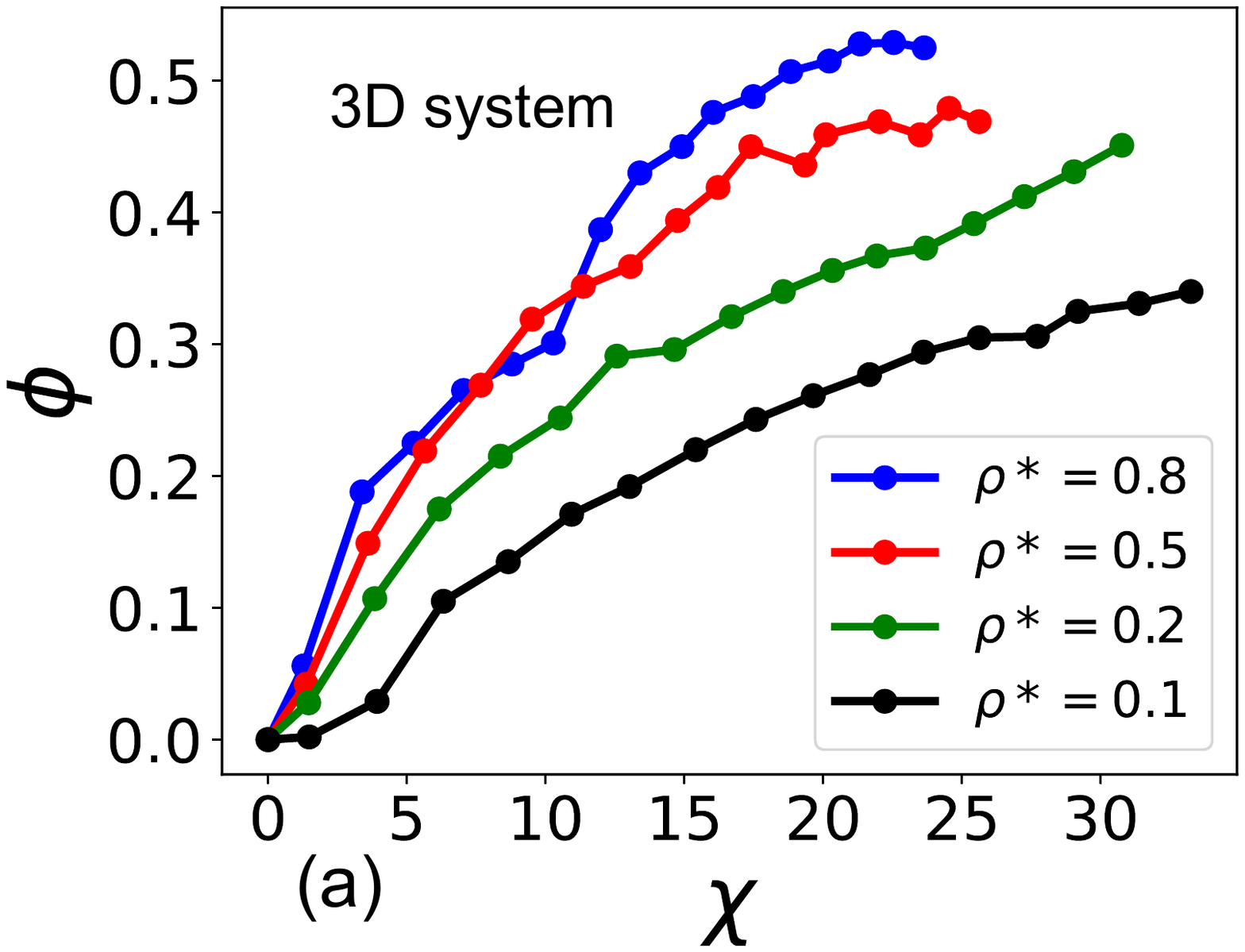}

}
\subfigure[]{
  \includegraphics[width=0.31\textwidth]{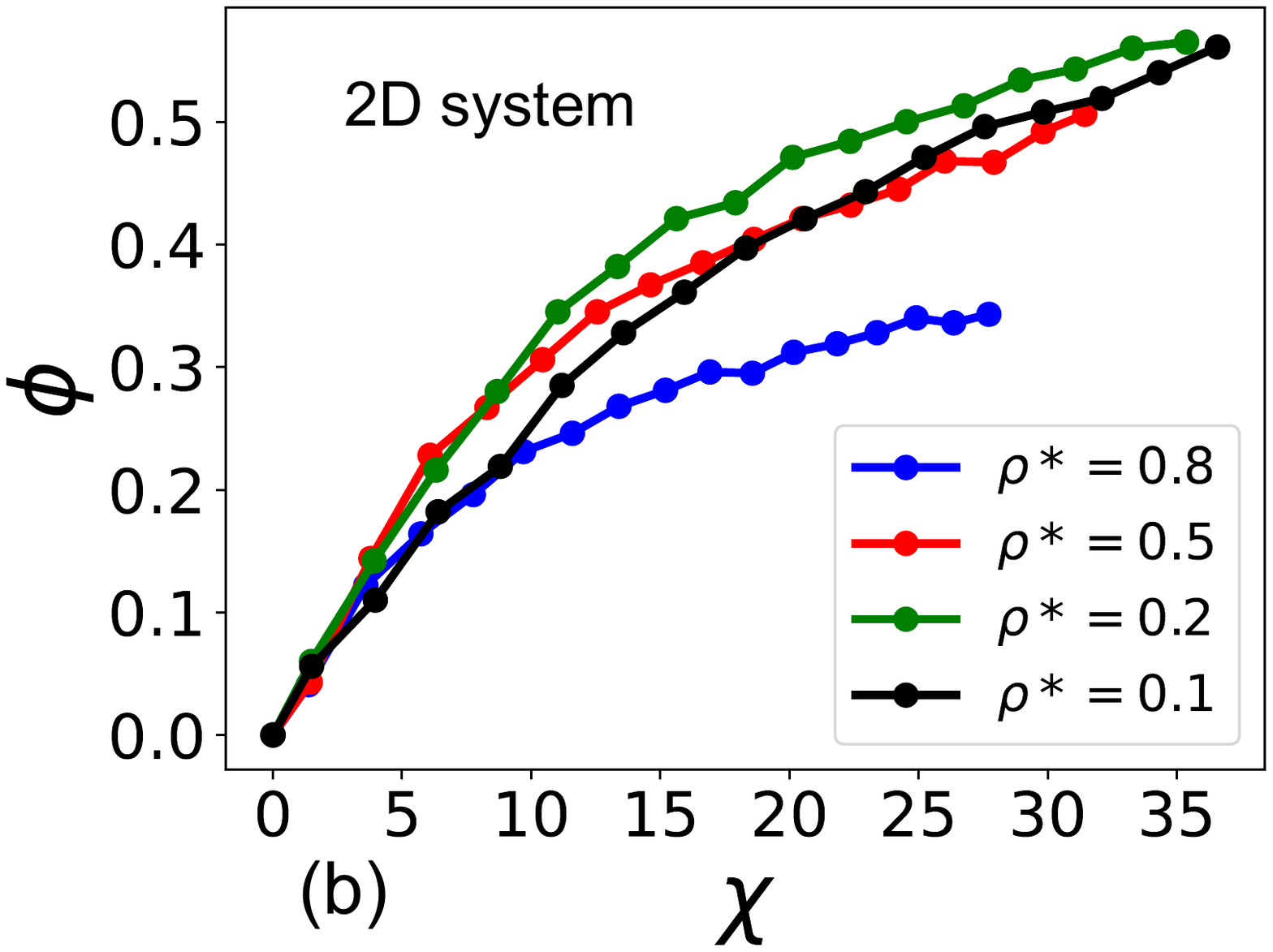}

}
\subfigure[]{
  \includegraphics[width=0.31\textwidth]{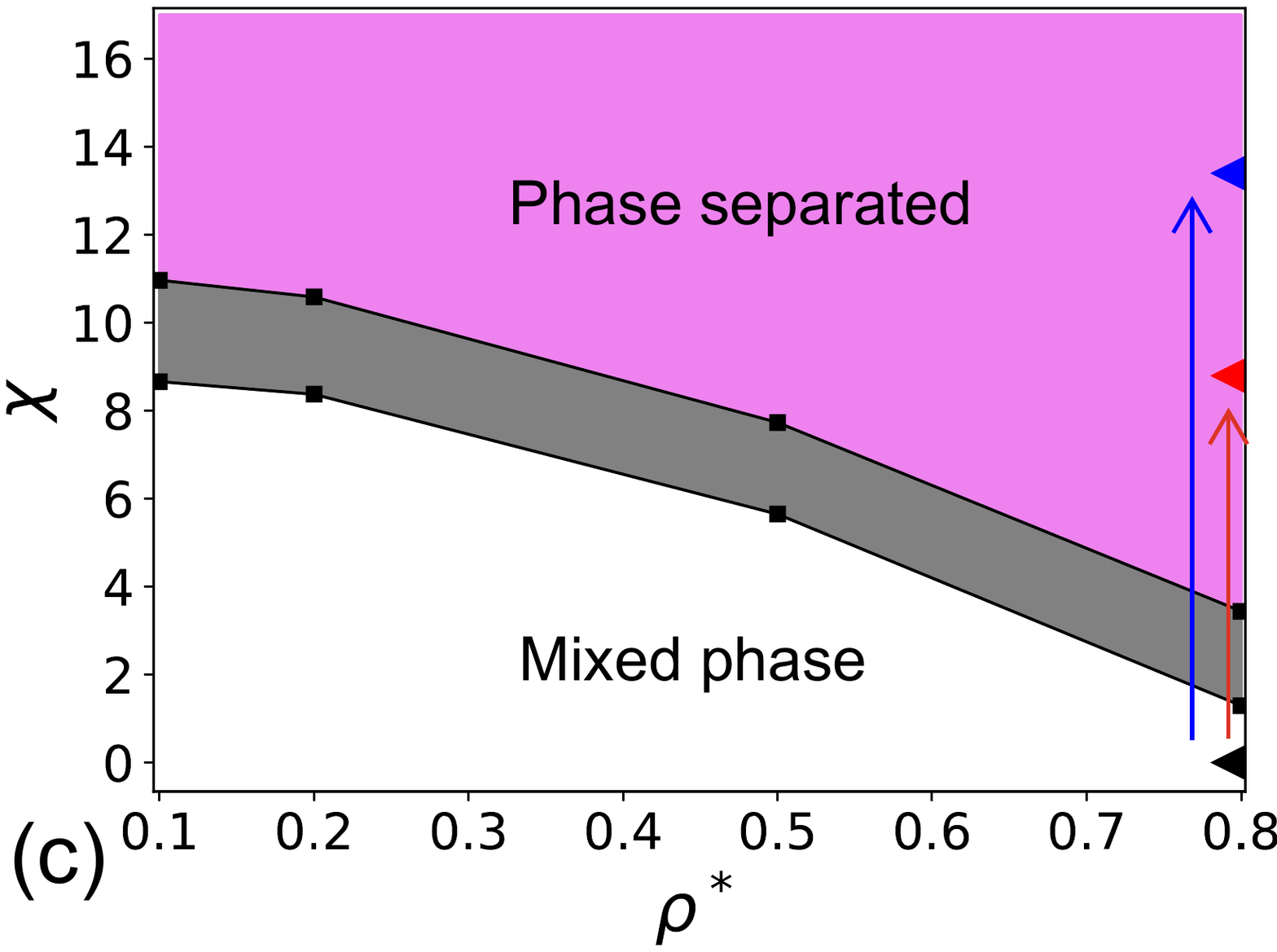}
}

\caption{(a) and (b) Plots of order parameter \(\phi\) versus activity \(\chi\) at different densities for 3D and 2D system respectively. The order parameter \(\phi\) increases with activity \(\chi\) at all densities. (c) Phase diagram of 2-TIPS in activity \(\chi\) and density \(\rho^*\) phase space for the 3D system. The purple color denotes the phase-separated region, and the white color indicates the mixed region of the phase space. The grey region is the critical region. The black triangle represents the initial state point corresponding to \(\rho^*=0.8, T_h^*=2, \chi=0\), which is quenched to the state points corresponding to \(\rho^*=0.8\), \(T_h^*=25\), \(\chi=8.79\) and \(\rho^*=0.8\), \(T_h^*=40\), \(\chi=13.39\) represented by red and blue triangles, respectively.  }
\label{fig:op_ss}
\end{figure*}
\FloatBarrier

\section{Snapshots for $3D$ system}\label{app:snap3d}

\setcounter{figure}{0}
\renewcommand{\thefigure}{B\arabic{figure}}

\FloatBarrier
\begin{figure*}[hbt]
    \centering
    \includegraphics[width=0.7\linewidth]{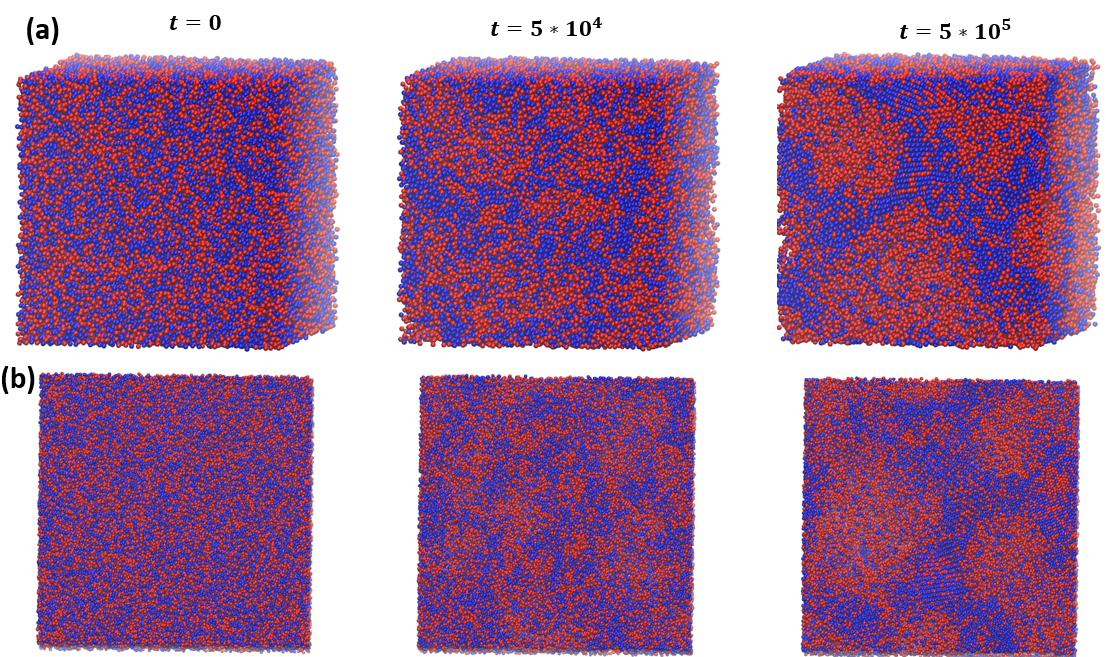}
    \caption{(a) The figures depict the instantaneous configurations of the 3D binary mixture of 80,000 hot(red) and cold(blue) particles at different instances of time following the quenching of the system to \(T_h^*=25\). (b) The figures show the top view of the simulation volume of the configurations given in Figure (a). The visuals show that the phase separation in 3D proceeds by the formation of bi-continuous domains with cold/hot particle-rich regions, which grow in size with time. }
    \label{fig:con3d}
\end{figure*}
\FloatBarrier

\FloatBarrier
\begin{figure*}[hbt]
    \centering
    \includegraphics[width=0.8\textwidth]{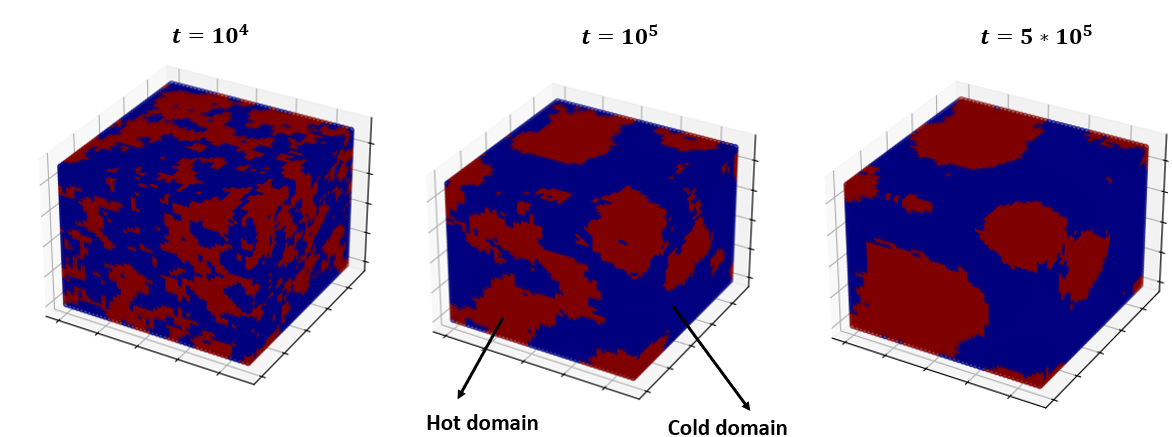}
    \caption{Order parameter field \(\phi(r,t)\) for quench temperature \(T_h^*=25\) and density \(\rho^*=0.8\) at various instants of time t for MD. The cold particle rich regions with high density (\(\rho^*(r)>0.8\)) are assigned \(\phi(r)=1\) and are depicted in dark blue color, while hot particle-rich regions with low density  (\(\rho^*(r)<=0.8\)) are assigned \(\phi(r)=-1\) and are represented by dark red color. The hot and cold regions form bi-continuous domains which grow in length with time t.}
    \label{fig:8}
\end{figure*}
\FloatBarrier

\begin{figure*}[htb]
    \centering
    \includegraphics[width=0.98\linewidth]{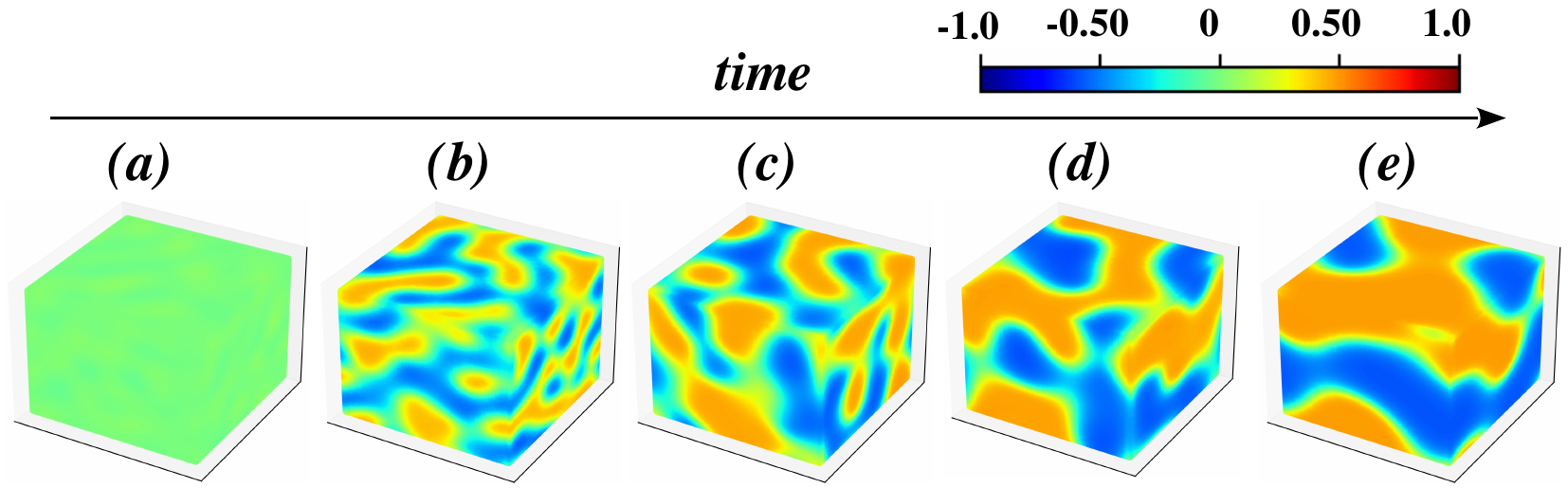}
    \caption{The figure shows the evolution of the system starting from a homogeneous initial state for CG System in 3D. The series of snapshots (a)-(e) depicts the plot of phase separation order parameter $\phi$ according to the color bar at subsequently increasing times : (a) $t = 10$, (b) $t = 50$, (c) $t = 150$, (d) $t = 300$, and (e) $1000$. parameters : Activity, $\chi = 2.00$; System size, $L = 32$. }
    \label{fig:critsnaps}
\end{figure*}
\FloatBarrier

\section{Snapshots for $2D$ system}\label{app:snap2d}

\setcounter{figure}{0}
\renewcommand{\thefigure}{C\arabic{figure}}

\FloatBarrier
\begin{figure*}[hbt]
\centering
  \includegraphics[width=0.9\textwidth]{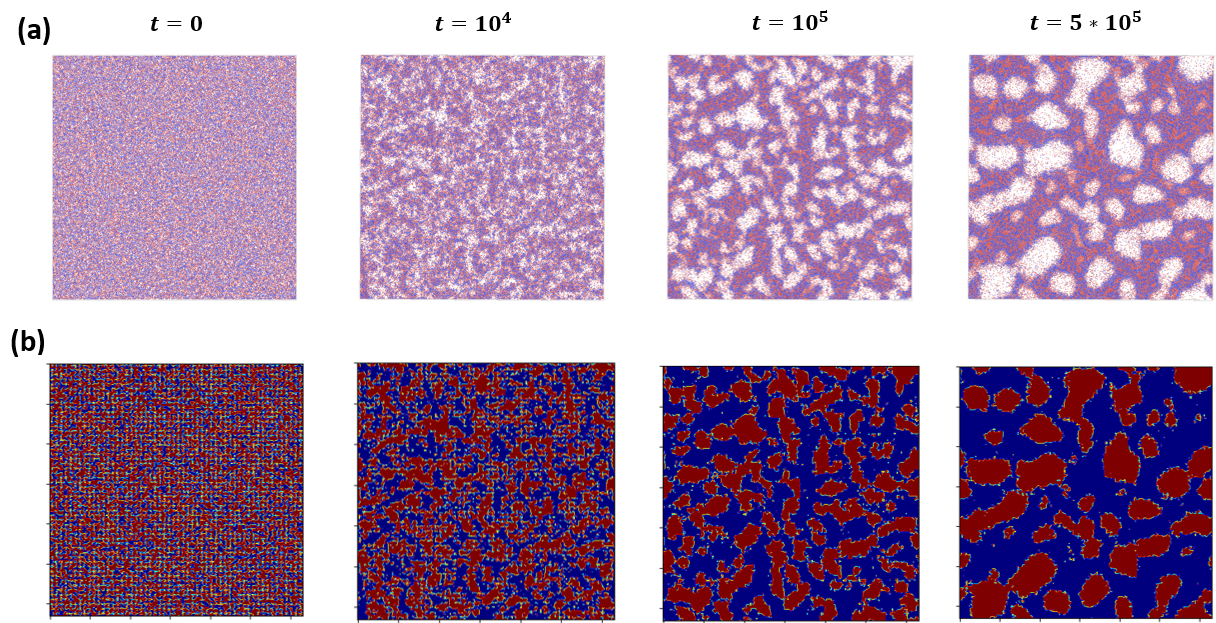}

\caption{ (a) Instantaneous configurations of 80,000 phase separating hot(red) and cold(blue) particles in 2D at various time intervals t, following the quenching of the system to \(T_h^*=25\). The figures illustrate the growth in bi-continuous domains enriched in either hot or cold particles over time. (b) The figures depict the order parameter field \(\phi(r,t)\) corresponding to the snapshots in (a), at different instants of time for quench temperature \(T_h^*=25\). The dark blue regions indicate domains with majority of cold particles (\(\phi(r)=1\)), and dark red regions correspond to domains rich in hot particles (\(\phi(r)=-1\)) }
\label{fig:11}
\end{figure*}
\FloatBarrier

\FloatBarrier
\begin{figure*}[htb]
   \centering
   \includegraphics[width=0.9\linewidth]{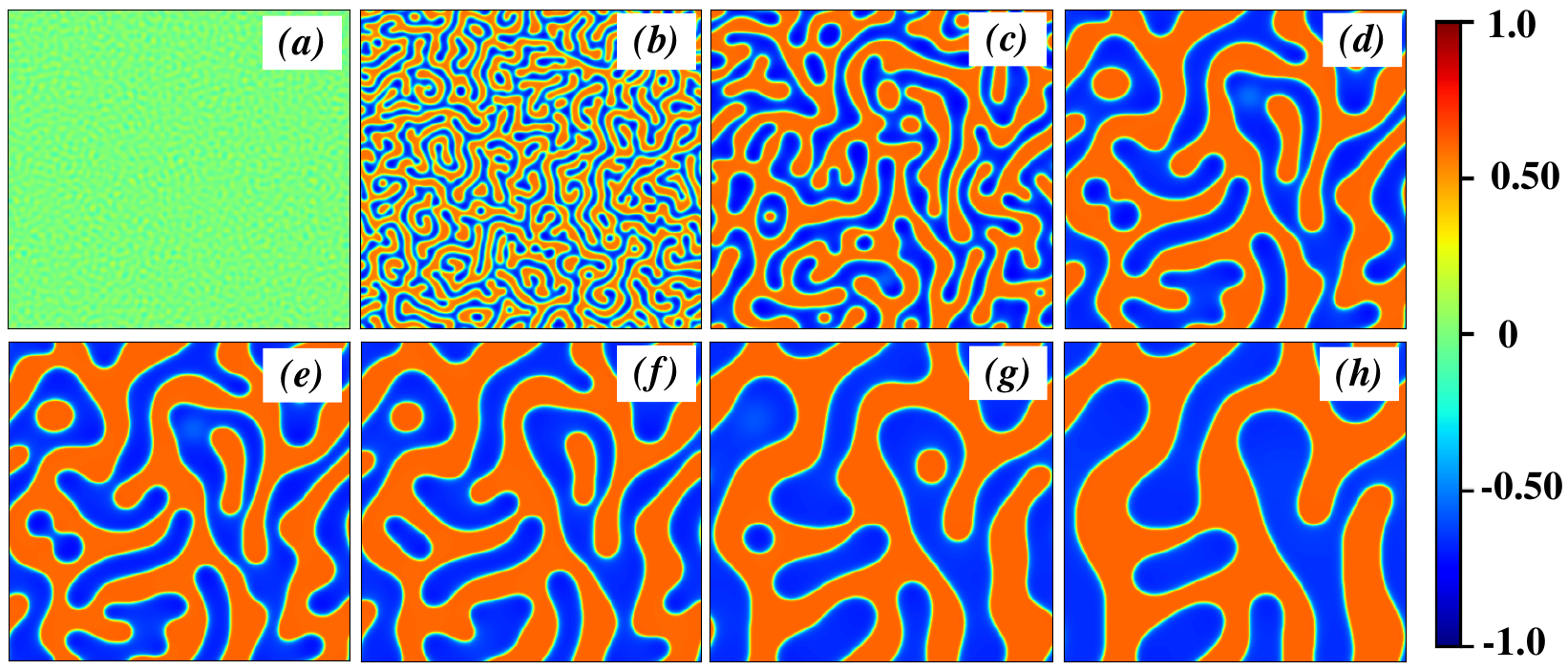}
   \caption{The figure shows the evolution of the 2D critical mixture starting from homogeneous state for a fixed activity. The snapshots (a-h) depict the configuration of $\phi$ according to the colorbar at subsequent times: (a) $5$, (b) $50$, (c) $5 \times 10^2$, (d) $2 \times 10^3$, (e) $2.4 \times 10^3$, (f) $3.6 \times 10^3$, (g) $5.2 \times 10^3$, and (g) $10^4$, respectively. In the snapshots, $\phi < 0$ and $\phi > 0$ represents regions dominated by cold and hot particles, respectively. System size, $L = 256$, and activity, $\chi = 2.50$. }
   \label{fig:2Dcrit_timeseries}
\end{figure*}
\FloatBarrier

\section{Current in the system}\label{app:current2d}
\setcounter{figure}{0}
\renewcommand{\thefigure}{D\arabic{figure}}

\begin{figure*}[hbt]
    \centering
    \includegraphics[width=0.7\linewidth]{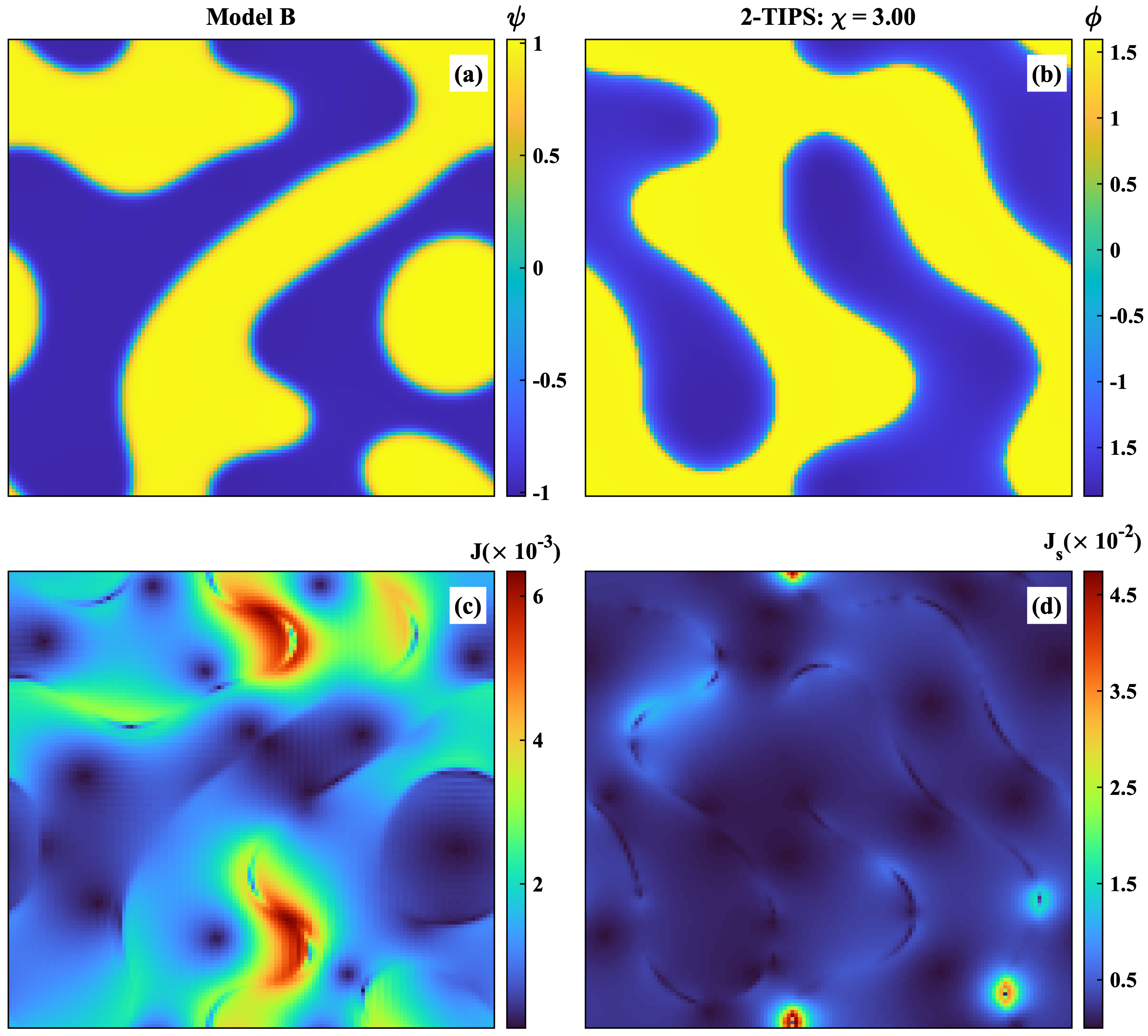}
    \caption{(Top row) The plot of the order parameter ($\psi(\boldsymbol{r})$) in subplot (a) and phase separation order parameter ($\phi(\boldsymbol{r}) = \psi_h(\boldsymbol{r}) - \psi_c(\boldsymbol{r})$) for 2-TIPS model in subplot (b). (bottom row) The plot shows the  the magnitude of current ($\vert \boldsymbol{J}(\boldsymbol{r}) \vert$) for model B in subplot (c) and the magnitude of current $\vert \boldsymbol{J}_{s}\vert = \vert \boldsymbol{J}_h + \boldsymbol{J}_c \vert$ for the 2-TIPS model in subplot (d) for activity $\chi = 3.00$. Parameters : System Size, L = 128 for 2D system. The rest of the parameters are same as in FIG.8 in the main text.}
    \label{fig:current_snap}
\end{figure*}

\begin{figure*}[hbt]
    \centering
    \includegraphics[width=0.5\linewidth]{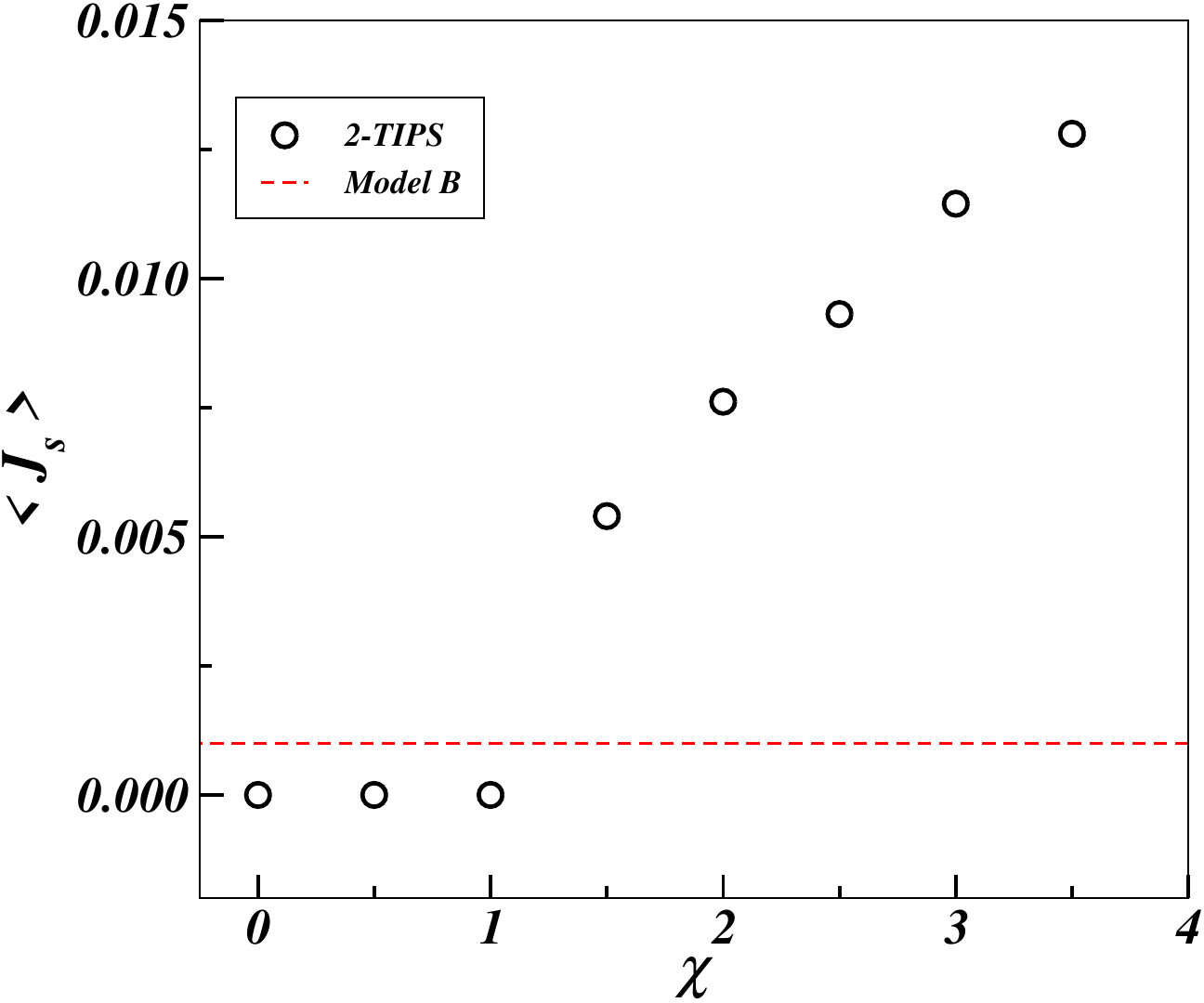}
    \caption{The figure shows the plot of $\langle J_{s}\rangle$ vs. activity $\chi$ for 2D system. The red dashed line shows the magnitude of current for model B. System Size : L=128.}
    \label{fig:current_mag}
\end{figure*}

We compare the current in the 2-TIPS system with model B in $two-$dimensions. The current for model B, $\boldsymbol{J}$, is given by Eq.3 in the main text. For the 2-TIPS model we calculate the total current, $\boldsymbol{J}_s = \boldsymbol{J}_c+ \boldsymbol{J}_h$, where, $\boldsymbol{J}_c = \boldsymbol{\nabla} \bigg(\frac{\delta F_{mix}}{\delta \psi_c} \bigg)$ and $\boldsymbol{J}_h = \boldsymbol{\nabla} \bigg(\frac{\delta F_{mix}}{\delta \psi_h} \bigg)$ are the current for the cold and hot species, respectively. In FIG.\ref{fig:current_snap}, we show the snapshot of current where the colorbar shows mgnitude of $\boldsymbol{J}$ (in subplot (c)) and $\boldsymbol{J}_s$ (in subplot (d)), respectively.\\
Further, we also show the plot the variation of mean magnitude of current $\boldsymbol{J}_s$, denoted by $\langle J_{s}\rangle$, with $\chi$ in FIG.\ref{fig:current_mag}, where $\langle J_{s}\rangle = \frac{1}{L^2}\mathlarger{\sum}_{\boldsymbol{r}} \vert \boldsymbol{J}_h(\boldsymbol{r})+\boldsymbol{J}_c(\boldsymbol{r})\vert$. The plot clearly shows that, for the values of $\chi$ for which the system phase separates into how and cold regimes,  $\langle J_s \rangle $ is much larger than that for model B $\langle J \rangle = \frac{1}{L^2}\mathlarger{\sum}_{\boldsymbol{r}} \vert \boldsymbol{J}(\boldsymbol{r}) \vert$. 
\FloatBarrier

\FloatBarrier
\FloatBarrier
\FloatBarrier
\twocolumngrid

\nocite{*}

\bibliography{apssamp}

\end{document}